\documentclass[a4paper,11pt]{article}
\pdfoutput=1
\usepackage{jheppub}
\usepackage{amsthm}
\usepackage{amsmath}
\usepackage{graphicx}
\usepackage{multirow}
\usepackage{tabularx}
\usepackage{makecell}
\usepackage{booktabs}
\usepackage{wrapfig}
\usepackage{subcaption}
\usepackage{slashed}
\usepackage{amssymb}
\usepackage{enumitem}
\usepackage{braket}
\usepackage{bbm}
\usepackage{bm}
\usepackage{physics}
\usepackage{xcolor}

\usepackage{caption}
\usepackage{slashed,cases,comment,bm,here,colortbl}
\usepackage{framed}
\definecolor{shadecolor}{rgb}{0.95,0.95,0.95}
\newenvironment{claim}{\begin{shaded}\noindent\ignorespaces}{\end{shaded}}

\usepackage{MnSymbol}
\usepackage{mathtools}

\usepackage{comment}

\DeclareMathOperator{\sign}{sign}

\def\beq{\begin{equation}}
\def\eeq{\end{equation}}

\newcommand{\bea}{\begin{eqnarray}}
\newcommand{\eea}{\end{eqnarray}}
\newcommand{\be}{\begin{equation}}
\newcommand{\ee}{\end{equation}}

\newcommand{\skipMG}[1]{}

\begin{document}

\title{Self-consistent computation of pair production from non-relativistic effective field theories in the Keldysh-Schwinger formalism}

\author{Tobias Binder, }

\author{Edward Wang}

\affiliation{\it Physik Department, James-Franck-Stra\ss e~1,
Technische Universit\"at M\"unchen,\\
D--85748 Garching, Germany}



\abstract{Sommerfeld-enhanced annihilation cross sections in the presence of nearly zero-energy bound states can become so large that perturbative partial-wave unitarity appears to be violated. Previous literature incorporated the short-distance annihilation potential self-consistently into the computation of the Schrödinger wave function at the origin, leading to the unitarization of the Sommerfeld effect in vacuum. We employ non-relativistic effective field theory methods and the Keldysh-Schwinger formalism to additionally include pair-creation effects in the self-consistent computation of four-point correlation functions, which renders the unitarization temperature dependent. Up to small thermal corrections in the non-relativistic and dilute regime of the pairs, we confirm the previous results based on the Schrödinger equation approach for scattering states in vacuum. For the first time, we analyze bound-state contributions beyond their leading decay via annihilation. Interestingly, our self-consistent computation of the four-point correlation function shows that bound states remain on-shell in their out-of-equilibrium decay, even though their spectral functions take the form of Breit–Wigner distributions due to finite decay widths. While this may appear paradoxical, it aligns with expectations from earlier results based on exact analytic solutions of the Kadanoff–Baym equations for a decaying elementary particle in a thermal environment.
}

\maketitle
\flushbottom

\section{Introduction}
\label{sec:intro}

Non-relativistic pairs interacting via a long-range potential can exhibit sizeable distortions of their wave function at short relative distances. In particular, for attractive interactions the wave function at the origin can be strongly enhanced, leading to an increase of short-distance annihilation rates. This effect is often referred to as the Sommerfeld enhancement~\cite{https://doi.org/10.1002/andp.19314030302} (or Sakharov enhancement~\cite{Sakharov:1991}). For $s$-wave and a single channel, it can generically be encapsulated in the simple relation:
\begin{align}
(\sigma v) = (\sigma v)_0 \, \big|\psi(\mathbf{r}=0)\big|^2 \,. \label{eq:SEcrossx}
\end{align}
The annihilation cross section factorizes into the tree-level annihilation cross section, $(\sigma v)_0$, and the scattering wave function of the pair at the origin, which encodes the effect of  potential interactions. For an attractive Coulomb potential only, $\big|\psi(0)\big|^2 \propto 1/v$ for relative velocities which satisfy $v\ll \alpha$ where $\alpha$ is a generic coupling constant. Such a velocity enhancement of the annihilation cross section through potential exchanges has triggered a lot of attention in the dark matter community.

For many dark matter (DM) candidates at the multi-TeV scale and above, even electroweak interactions with much lighter Standard Model states induce sizable long-range potential effects between non-relativistic DM particles. These include Sommerfeld enhancement~\cite{Hisano:2003ec, Hisano:2006nn} as well as the existence of metastable bound states~\cite{March-Russell:2008klu, Shepherd:2009sa, Pospelov:2008jd, vonHarling:2014kha}. Both modify relic abundance predictions relative to tree-level cross section estimates and lead to distinctive indirect-detection signatures. Prime examples are Wino, Higgsino, and minimal DM candidates~\cite{Beneke:2019gtg, Beneke:2020vff, Mitridate:2017izz, Baumgart:2023pwn, Baumgart:2025dov, Aghaie:2025iyn,Griffith:2026hdi}, where numerical solutions of multi-dimensional Schrödinger equations with matrix potentials and coupled channels are required due to nearly degenerate coannihilating states. But also simpler single-channel dark-sector models with light mediators~\cite{vandenAarssen:2012vpm, Cirelli:2016rnw, Binder:2017lkj, Biondini:2021ccr}, often motivated by structure formation, feature such effects.

It is well-known that the solution of the Schrödinger equation with an attractive Yukawa potential can feature near-zero-energy $s$-wave bound states ($\ell=0$)~\cite{Hisano:2003ec}. For such cases, and a single channel, the squared scattering wave function at the origin has a $1/v^2$ scaling for small velocities, which is much stronger than the aforementioned $1/v$ scaling for Coulomb potentials. Consequently, the Sommerfeld-enhanced annihilation cross section in Eq.~\eqref{eq:SEcrossx} for a Yukawa potential can grow so large that it violates parametrically the upper bound on the $s$-wave inelastic cross section~\cite{Griest:1989wd},
\begin{align}
 (\sigma v) \leq (\sigma v)_{\text{uni}} \equiv \frac{4\pi}{m^2 v} \,,\label{eq:unibound}
\end{align}
where the inequality follows from the requirement of a unitary $S$-matrix. The origin of violation of this inequality can be understood from the way annihilation is usually treated in the presence of long-range potential interactions. Namely, Eq.~\eqref{eq:SEcrossx} is usually derived from the expectation value of a short-distance absorptive interaction,
\begin{align}
(\sigma v) \propto
\bra{\psi} \Im \delta V \ket{\psi} \approx \bra{\psi_0} \Im \delta V \ket{\psi_0} \;,
\end{align}
where $\Im  \delta V \propto m^{-2} \delta^3(\mathbf{r})$ encodes short-distance annihilation for $s$-wave. The approximate symbol treats $\Im  \delta V$ as a perturbation (Born approximation) and $\psi_0$ denotes the scattering wave function computed from the Schr\"odinger equation without this contribution. Such perturbative treatment works for the Coulomb potential. However, for the Yukawa potential, the enhancement of $\psi_0(0)$ from zero-energy bound states is so strong that a perturbative treatment breaks down and naive $1/m^2$ suppressed power counting in the short-distance operator, $\Im  \delta V$, fails. 

A natural way to address this problem is to compute $\bra{\psi} \Im \delta V \ket{\psi}$ \emph{self-consistently}~\cite{Blum:2016nrz}, i.e., $\psi$ is the solution of the Schr\"odinger equation including $\Im \delta V$ non-perturbatively. This can be done by using the standard Lippmann-Schwinger procedure. The resulting self-consistently computed annihilation cross section respects the inequality~\eqref{eq:unibound}. This restoration of partial-wave unitarity, or unitarization of the Sommerfeld effect, has been extended to higher partial waves, coupled channels and to other type of reactions~\cite{Flores:2024sfy,Parikh:2024mwa,Flores:2025uoh}, relevant also for the regulation of the excessive annihilation enhancement from quasi-bound states present for $\ell \geq 1$~\cite{Kamada:2023iol,Beneke:2024iev}.

In this work, we revisit the self-consistent computation of Sommerfeld-enhanced annihilation for the simplest case of s-wave and a single channel. While the previous Schr\"odinger equation approach~\cite{Blum:2016nrz} is clearly a sufficient description to unitarize Sommerfeld-enhanced annihilation rates for DM indirect detection predictions, our main motivation is to study possible effects from the reverse process, i.e. pair creation, which is relevant during early stages of the freeze-out process to achieve chemical equilibration of DM in the early Universe. Moreover, previous literature focused on scattering states. In this work, we extend the analysis by also self-consistently computing the decay of bound states via annihilation, which is mediated by the same operator $\Im \delta V$. This is motivated by the fact that bound states close to zero energy are tightly connected to the behavior of the enhancement of the annihilating scattering states. In the following, we refer to a self-consistent computation of \emph{pair} annihilation and creation to both, scattering and bound states.

Including temperature-dependent pair creation in the self-consistent solution necessitates a non-equilibrium quantum field theory approach. In particular we will work in the Keldysh-Schwinger formalism~\cite{PhysRevA.4.739, Keldysh:1964ud}. As the pairs are assumed to be non-relativistic, we can simplify the problem from the beginning by making use of non-relativistic (NR) effective field theories (EFTs). Combinations of NR EFTs and the Keldysh-Schwinger formalism have been used in the DM literature before, see Refs.~\cite{Binder:2018znk, Binder:2020efn, Binder:2021kse, Binder:2021otw, Biondini:2023zcz, Biondini:2024vmz, Biondini:2025jvp} (and \cite{Beneke:2014gla}, as well as Keldysh-Schwinger approaches also for freeze-in \cite{Becker:2023vwd,Becker:2025lkc} for complementarity). The theoretical framework is in principle established, however, a self-consistent computation of pair creation and annihilation in the presence of nearly zero-energy bound states is still unexplored. Moreover, it is unclear if bound states are on- or off-shell when chemically equilibrating, due to their finite decay width induced by annihilation.

We investigate this system, based on a static potential and pair annihilation/creation operator, from first principles. The procedure is the following. We use \emph{non-relativistic effective field theory} (NREFT)~\cite{Caswell:1985ui} techniques in the Keldysh-Schwinger formalism to derive a differential equation for the particle number density evolution. This transport equation will depend on a \emph{four-point correlation function} because of pair annihilation. A derivation has been presented in Ref.~\cite{Binder:2018znk} for NRQED at finite temperature, and we review the most relevant results for this work in Section~\ref{sec:numberNR}. Building on this, we extend Ref.~\cite{Binder:2018znk} by including the reverse process in the number density equation and self-consistently computing the four-point correlation functions, necessary for resonant phenomena. 

We perform the self-consistent computation of the four-point correlation function in two ways: i) the NREFT approach requires truncating the in principle infinitely coupled Martin-Schwinger correlation function hierarchy at the four-point correlation function level to get closed equations that we solve, and ii) using \emph{potential non-relativistic effective field theory} (pNREFT)~\cite{Pineda:1997bj, Brambilla:1999xf} techniques in the Keldysh-Schwinger formalism, which provides closed equations for the four-point correlation function at the two-particle sub-space level. In all computations we neglect, for simplicity, the contribution of ultrasoft fields that connect scattering and bound states. It is shown that both approaches deliver consistent results based on a static potential plus pair annihilation and creation.

The remainder of this work is organized as follows. In Section~\ref{sec:numberNR} we review the derivation of the particle number density equation from NREFT on the Keldysh-Schwinger contour and identify relevant four-point correlation functions. In Sections~\ref{sec:spectralfunction} and~\ref{sec:NRpaircreation}, we compute these self-consistently within the NREFT approach, whereas in Section~\ref{sec:pNREFTapproach} we perform the calculation within the pNREFT approach. By the end of Section~\ref{sec:NRpaircreation}, we also give an explicit example of how to match a model in practice. Technical details are shared in three Appendices. We conclude in Section~\ref{sec:conclusion}.

\section{Number density equation from NREFT}
\label{sec:numberNR}

We review the derivation of a differential equation for the particle number density from non-relativistic effective field theory (NREFT) at finite temperature, which has been given in Ref.~\cite{Binder:2018znk} for NRQED in the Keldysh-Schwinger formalism. While here we focus on a scalar particle and anti-particles, the essential parts of the derivation are the same and we refer to Ref.~\cite{Binder:2018znk} for details. 

The basic idea is to derive from NREFT in the Keldysh-Schwinger formalism a differential equation for the particle number density, that includes the effect of annihilation. The result is Eq.~\eqref{eq:number}, which shows that the right hand side depends on the s-wave tree-level annihilation cross section, which multiplies a four-point correlation function, $G_{\eta \xi}$, on the Keldysh-Schwinger contour. For the rest of the work, we solve this four-point correlation function under certain limits, where ultimately in subsequent Sections~\ref{sec:pNREFTapproach} and~\ref{sec:NRpaircreation}, we will solve it self-consistently. That means taking also annihilation in the solution of this four-point correlation function into account, which is of higher order, but relevant in the presence of zero-energy bound states as parametrically possible for Yukawa potential interactions.

\subsection{NREFT on CTP contour}
We consider a non-relativistic effective action on the closed-time path (CTP) contour for a scalar particle ($\eta$) and anti-particle ($\xi$) field with the same mass $m$. We assume that those can interact via a static potential, and particle-antiparticle pairs can annihilate into light degrees of freedom. The corresponding NR effective action on the CTP contour reads: 
\begin{align}
S_\text{NR} &= \int\limits_{x^0\in \mathcal{C}} \text{d}^4x \; \eta^\dagger \left( i\partial_{x^0} + \frac{\Delta}{2m}\right) \eta +\xi^\dagger\left( i\partial_{x^0} + \frac{\Delta}{2m}\right) \xi \nonumber \\ &+ \int\limits_{x^0,y^0 \in \mathcal{C}}\text{d}^4x\text{d}^4y \; \eta^\dagger(x) \xi(x) i \Gamma(x,y) \xi^\dagger(y) \eta(y) + \text{static potential} \;.\label{eq:NRaction}
\end{align}
where $x^0,y^0$ are on the CTP contour denoted as $\mathcal{C}$. Annihilation effects are encoded in the ``matching coefficient" $\Gamma$ after integrating out the light degrees of freedom at one-loop order. Assuming s-wave or contact annihilation, it has the following \emph{matrix structure} on the CTP contour (see Appendix B in \cite{Binder:2018znk} for a generic case, or Appendix~\ref{app:example} here for a concrete scalar example):
\begin{align}
\Gamma(x,y)= \frac{(\sigma v)_0}{2} \delta^4(x-y)
\begin{pmatrix}
     1& 0\\
    2  & 1
\end{pmatrix} \;. \label{eq:annihop}
\end{align}
Here, $(\sigma v)_0$ denotes the constant s-wave tree-level annihilation cross section. 
Notice that we have not yet included the effect of the reverse process, i.e. pair creation, considered later in Sections~\ref{sec:exact} and \ref{sec:NRpaircreation}. It leads to a non-local and temperature dependent form of $\Gamma(x,y)$. For now, we will consider only annihilation and assume that there are no light degrees of freedom in the environment that can pair produce the heavy particle and anti-particle fields. In the DM cosmological context, this is justified for temperatures much smaller than the typical freeze-out temperature $T=m/25$ where pair creation effects can be neglected as they are exponentially suppressed. To achieve chemical equilibration (at early times for DM), these exponentially suppressed terms become relevant, and one has to include them in the finite T EFT which requires a careful power counting beyond the standard vacuum treatment of NREFT (see Appendix~\ref{app:example}), and is illustrated in Section~\ref{sec:NRpaircreation}.

Eq.~\eqref{eq:NRaction} can be seen as a phenomenological model to study pair annihilation in the presence of a static potential.
The only assumption we will make on the static potential part is that the potential interaction is particle and anti-particle number conserving, so that it will drop out of the particle number density Eq.~\eqref{eq:number}, see Ref.~\cite{Binder:2018znk} for an explicit example in QED.

Notice that the kinetic terms do not involve, e.g. dynamical gauge field interactions, and therefore we miss the contribution of ultrasoft fields that connect bound and scattering states. Consequently, in this toy model, scattering and bound states will evolve independently, which is sufficient for us as we are only interested in higher order annihilation effects in a simplified picture without ultrasoft fields. Including the ultrasoft fields self-consistently may unitarize, e.g., bound-state formation, an interesting topic on its own~\cite{Oncala:2019yvj,Binder:2023ckj,Beneke:2024nxh,Flores:2024sfy,Flores:2025uoh,Flores:2026yay}.

\subsection{Number density equation: coupled two- and four-point functions}
From the NREFT action on the CTP contour in Eq.~\eqref{eq:NRaction}, we derive the equations of motion of the particle two-point correlation function, which depends on four-point correlation functions due to annihilation and static potential interactions. Taking the equal spacetime limit in the kinetic equation, we get for the particle number density evolution equation~\footnote{Throughout this work we assume the system to be spatially homogeneous and isotropic. In particular, this implies that the four-point correlator $G^{++--}_{\eta \xi}(x,y,z,w)$ does not depend on the Wigner coordinate $\mathbf{X}=(\mathbf{x}+\mathbf{y}+\mathbf{z}+\mathbf{w})/4$, see Appendix~\ref{app:NREFT}. In cosmological settings, one has to replace the time derivative by $\partial_T \rightarrow \partial_T + 3 H$ in Eq.~\eqref{eq:number}, where $H$ is the (Hubble) expansion rate of the Universe.}:
\begin{claim}
\begin{align}
\dot n_\eta = -  (\sigma v)_0 G^{++--}_{\eta \xi}(x,x,x,x) \label{eq:number} \;.
\end{align}
\end{claim}
The static potential part canceled out as it is number conserving. The particle number density, $n_\eta$, is defined from an interacting two-point correlation function at equal spacetime coordinates as:
\begin{align}
n_\eta \equiv \langle \eta^\dagger(x) \eta (x) \rangle \,,
\end{align}
where $\eta^\dagger \eta$ is the particle number density operator. The four-point correlation function is defined as:
\begin{align}
G_{\eta \xi}(x,y,z,w) \equiv \langle T_\mathcal{C} \eta(x) \xi^\dagger(y)  \xi(w) \eta^\dagger(z) \rangle \;,\label{eq:def4point}
\end{align}
where $T_\mathcal{C}$ denotes time ordering on the CTP contour $\mathcal{C}$. From the NREFT equation of motion of this four-point correlation function, one can show that at leading order the four-point function factorizes into uncorrelated number densities as:
\begin{align}
G_{\eta \xi}^{++--}(x,x,x,x) \simeq G_\eta^{+-}(x,x) G^{+-}_\xi(x,x) = n_\eta n_\xi \,.
\end{align}
Inserting this into \eqref{eq:number} gives a quadratic number density equation in a symmetric plasma where $n_\eta= n_\xi$. It is consistent with expectations from the standard Boltzmann equation where particles and anti-particles are uncorrelated, and pair creation can be neglected. 

Static potential insertions in the solution of the four-point function introduce a correlation, which is strictly speaking beyond this factorized form. Namely, the more general four-point correlation function can feature bound states for sufficiently long-ranged potentials, leading to a linear dependence on the bound state number density. In such a case, $n_\eta$ on the left hand side of Eq.~\eqref{eq:number} also counts the particle number inside the particle-antiparticle bound states. Therefore, $n_\eta$ has to be interpreted as the total particle number density.

One can understand this in an intuitive way. The left hand side of Eq.~\eqref{eq:number} is defined in terms of an interacting two-point correlation function, thus being exact. The right hand side, in practice, can never be evaluated exactly (except for lattice simulations). Thus, depending on the approximations of the right hand side involving the four-point correlation function, one has to adapt the interpretation of the left hand side accordingly. We turn to a concrete example with scattering and bound states in the next section.

\subsection{KMS relation}
\label{sec:KMS}

At the next level of computing the four-point correlation function, we aim to demonstrate explicitly that it encodes both the Sommerfeld enhancement and possible bound-state contributions. If bound states exist, the number density $n_\eta$ appearing on the left-hand side of Eq.~\eqref{eq:number} must be interpreted as the \emph{total} particle number density, including both scattering states and bound states.

To make this explicit, let us start with the case where we compute the four-point correlation function in the absence of annihilation, which works for static potentials in parameter regions where no zero-energy bound states can appear. To do so, we define a four-point correlation function without annihilation, by formally switching off the annihilation width $\Gamma$,
\begin{align}
G_{\eta\xi,0}(x,y,z,w) \equiv G_{\eta\xi}(x,y,z,w)\big|_{\Gamma=0} \;.
\label{eq:defgppmmfree}
\end{align}
In this limit, particle and antiparticle numbers are separately conserved in the solution of $G_{\eta\xi,0}$. This follows from the presence of two independent global $U(1)$ symmetries of the non-relativistic (NR) Lagrangian in Eq.~\eqref{eq:NRaction} when annihilation is neglected. Equivalently, the corresponding number operators $\hat{N}_\eta$ and $\hat{N}_\xi$ commute with the NR Hamiltonian without annihilation, $\hat{H}_{\mathrm{NR},0}$. As a consequence, independent chemical potentials can be consistently associated with the conserved charges $\hat{N}_\eta$ and $\hat{N}_\xi$. This allows us to describe the system by a grand-canonical density matrix,
\begin{align}
\hat{\rho}
= \exp\!\left[
-\beta\!\left(
\hat{H}_{\mathrm{NR},0}
+ m(\hat{N}_\eta + \hat{N}_\xi)
- \mu_\eta \hat{N}_\eta
- \mu_\xi \hat{N}_\xi
\right)
\right] \;,
\end{align}
where $\mu_\eta$ and $\mu_\xi$ are the chemical potentials. In the following, we restrict ourselves to a symmetric plasma and set $\mu \equiv \mu_\eta = \mu_\xi$. In this grand-canonical ensemble, the Kubo-Martin-Schwinger (KMS) relation can be used to express the four-point correlator $G^{++--}_{\eta\xi,0}$ in terms of the corresponding spectral function $G^\rho_{\eta\xi,0}$,
\begin{align}
G_{\eta\xi,0}^{++--}(x,x,x,x)
&= \int \frac{\mathrm{d}\omega}{(2\pi)} \frac{\mathrm{d}^3 P}{(2\pi)^3}
G_{\eta\xi,0}^{++--}\!\left(T,\omega,\mathbf{P},\mathbf{r}=0,\mathbf{r}^\prime=0\right)
\nonumber \\
&= \int \frac{\mathrm{d}\omega}{(2\pi)} \frac{\mathrm{d}^3 P}{(2\pi)^3}
f_B^{\mathrm{eq}}\!\left(2m + \omega - 2\mu \right)
G^\rho_{\eta\xi,0}\!\left(\omega,\mathbf{P},\mathbf{r}=0,\mathbf{r}^\prime=0\right)
\nonumber \\
&\simeq \int \frac{\mathrm{d}\omega}{(2\pi)} \frac{\mathrm{d}^3 P}{(2\pi)^3}
e^{-\beta (2m + \omega - 2\mu)}
G^\rho_{\eta\xi,0}\!\left(\omega,\mathbf{P},\mathbf{r}=0,\mathbf{r}^\prime=0\right)
\nonumber \\
&= e^{2\beta\mu}
\int \frac{\mathrm{d}^3 P}{(2\pi)^3}
e^{-\beta (2m + \mathbf{P}^2/(4m))}
\int_{-\infty}^{+\infty} \frac{\mathrm{d}E}{(2\pi)}
e^{-\beta E}
G^\rho_{\eta\xi,0}\!\left(E;\mathbf{r}=0,\mathbf{r}^\prime=0\right) \;.
\label{eq:spectral}
\end{align}
In the first line, we have switched to Wigner coordinates, see Appendix~\ref{app:NREFT} for convention. In the second line, the KMS relation has been applied, while in the third line, we have approximated the Bose-Einstein distribution $f_B^{\mathrm{eq}}$ by a Maxwell-Boltzmann distribution, valid in the dilute limit. In the last line, we introduced the variable $E \equiv \omega - \mathbf{P}^2/(4m)$, as the Wigner time $T$ independent spectral function depends only on this combination of $\omega$ and the center-of-mass kinetic energy $\mathbf{P}^2/(4m)$.

By closing the Martin-Schwinger hierarchy at the four-point function level, one can obtain the solution of the spectral function from the retarded Green's function, which fulfills the equation in the dilute limit:
\begin{align}
\bigg[ E + i \epsilon + \frac{\Delta_\mathbf{r}}{m} - V(r) \bigg]G^R_{\eta \xi,0}(E;\mathbf{r},\mathbf{r}^\prime) = i \delta(\mathbf{r}-\mathbf{r}^\prime) \;.
\end{align}
In spectral representation, the solution of this equation is given by:
\begin{align}
G^R_{\eta \xi,0}(E;\mathbf{r},\mathbf{r}^\prime) = i \sum_{\mathcal{B}}\frac{\psi_{\mathcal{B}}(\mathbf{r}) \psi^\star_{\mathcal{B}}(\mathbf{r}^\prime)}{E-E_{\mathcal{B}} + i \epsilon} + i \int \frac{\text{d}^3q}{(2\pi)^3} \frac{\psi_\mathbf{q}(\mathbf{r}) \psi^\star_\mathbf{q}(\mathbf{r}^\prime)}{E-q^2/m + i \epsilon} \,,
\end{align}
from which one can obtain the spectral function (retarded minus advanced) at the origin as:
\begin{align}
G^\rho_{\eta \xi,0}(E;0,0) = 2 \Im i G^R_{\eta \xi,0}(E;0,0) = \sum_{n} |\psi_{n}(0)|^2 (2\pi)\delta(E-E_n) + \theta(E) \frac{m^2}{2\pi} \sqrt{\frac{E}{m}} S(v) \;, \label{eq:Grhoetaxi0}
\end{align}
where summation over the principal quantum nummber, $n$, includes all existing $s$-wave bound states, $S(v)=|\psi_\mathbf{p}(0)|^2$ is the Sommerfeld factor for kinetic energies $E=p^2/m=mv^2/4$, and only $s$-wave scattering states contribute at the origin. Inserting Eq.~\eqref{eq:Grhoetaxi0} into Eq.~\eqref{eq:spectral}, the resulting number density equation contains the (unregulated) Sommerfeld effect and bound state decay, and is consistent with standard Boltzmann equations in the limit of ionization equilibrium \cite{Binder:2018znk, Binder:2019ikc}\footnote{In the presence of bound states, the chemical potential $\mu$ in \eqref{eq:spectral} is a non-trivial function of the total particle number $n_\eta$ that appears on the left hand side in \eqref{eq:number}. Describing the system with only one chemical potential amounts to assume that the system is in ionization equilibrium, which sets a relation between single particle and bound state number densities. This is the limit where ultrasoft transitions are efficient.}.

So far, we have neglected the effect of annihilation in the solution of $G^{++--}_{\eta \xi,0}$. Importantly, in the presence of a Yukawa potential, the wave function at the origin for scattering states can become so large, that $(\sigma v)_0 \times S(v) $ in Eq.~\eqref{eq:number} can parametrically violate the partial wave unitarity bound for inelastic $2 \rightarrow 2$ collisions, which was introduced in Eq.~\eqref{eq:unibound}. We turn to a self-consistent computation of $G^{++--}_{\eta \xi}$, including the effect of annihilation in its solution.

\section{Self-consistent spectral function}
\label{sec:spectralfunction}
The EoM of the retarded four-point correlation function, including annihilation, is given by:
\begin{align}
\bigg[ E + i \epsilon + \frac{\Delta_\mathbf{r}}{m} - V(r) + i c \delta(\mathbf{r})\bigg]G^R_{\eta \xi}(E;\mathbf{r},\mathbf{r}^\prime) = i \delta(\mathbf{r}-\mathbf{r}^\prime) \;,
\end{align}
where we introduced $c\equiv (\sigma v)_0/2$ for notational simplicity. Perturbing around the solution without annihilation, the full retarded Green's function fulfills the equation:
\begin{align}
G^R_{\eta \xi}(E;\mathbf{r},\mathbf{r}^\prime) &= G^R_{\eta \xi,0}(E;\mathbf{r},\mathbf{r}^\prime) - \int \text{d}^3 \bar{r} G_{\eta \xi,0}^R(E;\mathbf{r},\bar{\mathbf{r}}) c \delta(\bar{\mathbf{r}}) G^R_{\eta \xi}(E;\bar{\mathbf{r}},\mathbf{r}^\prime) \nonumber \\
&= G^R_{\eta \xi,0}(E;\mathbf{r},\mathbf{r}^\prime) - cG_{\eta \xi,0}^R(E;\mathbf{r},0) G^R_{\eta \xi}(E;0,\mathbf{r}^\prime) \;.
\end{align}
This can be solved, and for the full retarded correlator at the origin the solution is:
\begin{align}
G^R_{\eta \xi}(E;0,0) = \frac{G^R_{\eta \xi,0}(E;0,0)}{1+c G^R_{\eta \xi,0}(E;0,0)} \;. \label{eq:selfconsistentretarded}
\end{align}
The imaginary part of the retarded Green’s function at the origin, $G^R_{\eta\xi,0}(E;0,0)$, is divergent. There exist different treatments of the divergence, see~\cite{Blum:2016nrz,Parikh:2024mwa}, and c.f.~\cite{Braaten:2017dwq},~\cite{Flores:2024sfy},~\cite{Watanabe:2025kgw}, and~\cite{Flores:2025uoh}. For a recent comparison, see \cite{Flores:2025uoh, Cimring:2026}.


Crucially, the same regularization scheme used to define matching coefficients (such as $c$) must also be employed in intermediate computations such as of the retarded Green’s function at the origin. The most commonly used scheme in non-relativistic EFT calculations is dimensional regularization, in which the retarded Green's function at the origin is defined as:
\begin{align}
G^{R}_{\eta\xi,0}(E;0,0) &= \int \frac{\text{d}^{d-1} p}{(2\pi)^{d-1} }  \int \frac{\text{d}^{d-1}  p^\prime}{(2\pi)^{d-1} } G^R_{\eta\xi,0}(E;\mathbf{p},\mathbf{p}^\prime)\;. 
\end{align}
Let us consider the zero and one potential exchange in momentum space:
\begin{align}
G^{R}_{\eta\xi,0}(E;\mathbf{p},\mathbf{p}^\prime) &= i\frac{(2\pi)^{d-1}\delta^{d-1}(\mathbf{p}-\mathbf{p}^\prime)}{E-p^2/m + i \epsilon} + i \frac{1}{E-p^2/m + i \epsilon} V(\mathbf{p}-\mathbf{p}^\prime) \frac{1}{E-p^{\prime 2}/m + i \epsilon}  +\mathcal{O}(V^2)\label{eq:coulombexch}
\end{align}
The zero exchange is linearly ultraviolet (UV) divergent. For a Coulomb potential, $V(r)=-\alpha/r$, the one exchange is logarithmically UV divergent. All higher Coulomb exchanges are finite, leading to the Coulomb retarded Green's function in dimensional regularization ($d=4-2\epsilon$)~\cite{Beneke:1999zr, Eiras:1999xx, Beneke:2013jia}:
\begin{align}
G^{R}_{\eta\xi,0}(E;0,0) &=i \frac{m^2}{4\pi} \left(  \sqrt{-\frac{(E+i \epsilon)}{m}} + \alpha \bigg[ \frac{-1}{4\epsilon}+ \frac{1}{2} \ln \left( \frac{-4m(E+i \epsilon)}{\mu^2} \right) - \frac{1}{2}  + \hat{\Psi}\left(1-\lambda \right) \bigg] \right) + \mathcal{O}(\epsilon) \label{eq:msbargrfct}
\end{align}
where $\lambda \equiv \alpha/(2\sqrt{-(E+i\epsilon)/m})$, and $\hat{\Psi}(x)=\gamma_E + \Psi(x)$, where $\Psi$ is the Euler Psi function $\Psi(x)=\frac{\text{d}}{\text{d}x} \ln \Gamma(x)$. Notice the presence of an uncanceled UV divergence, $1/\epsilon$, and a renormalization-scale dependence on $\mu$. Since the finite mediator mass in the Yukawa potential does not modify the ultraviolet behavior, we likewise expect a divergent one–potential-exchange contribution, while higher-order exchanges remain finite. Such uncanceled UV divergences and $\mu$ dependencies are not unexpected, as we work with a restricted subset of matching coefficients and operators in NREFT (and later pNREFT), retaining only the annihilation operator and its unspecified matching coefficient $c$.

Consider top-quark pair production near threshold as a concrete theory where a spectral function at the origin is computed, including perturbative corrections from delta-potential insertions. As discussed in Sections~2.5.4 and 2.5.5 of Ref.~\cite{Beneke:2024sfa}, when all operators, matching coefficients, and coupling runnings are included, both $1/\epsilon$ poles and $\mu$ dependencies cancel up to the expected order in accuracy. Notice that in the dilute limit, our retarded and the time-ordered non-relativistic Green’s functions adopted in vacuum (top-quark pair production) coincide, yielding identical spectral functions.

In our simplified setup, where the annihilation operator with an open matching coefficient 
 $c$ is effectively resummed to all orders while other operators, are neglected, residual 
$\log\mu$ dependence can arise. Determining which explicit form of $c$, or potentially required additional matching coefficients or operators cancel this dependence lies beyond the scope of this work.\footnote{In general, further $\mu$-dependent and in general complex contributions multiplying the delta potential are present; see Ref.~\cite{Pineda:1998kn}. In complete calculations of observables, such as the Lamb shift~\cite{Pineda:1997ie}, $\mu$ dependence cancels once all contributions (including ultrasoft) are consistently included.} We therefore remove the $1/\epsilon$ divergence minimally and proceed by treating $G^R_{\eta\xi,0}(E;0,0)$ as having a finite (but $\mu$-dependent) imaginary part. In this way we leave, given the current literature, other treatments such as ~\cite{Blum:2016nrz,Parikh:2024mwa} or \cite{Flores:2025uoh} as potential ways to continue the renormalization procedure.\footnote{Following Ref.~\cite{Blum:2016nrz} and subtracting the imaginary part of Eq.~\eqref{eq:msbargrfct} at a reference energy $E_0>0$ reproduces the Coulomb example in Ref.~\cite{Blum:2016nrz},
\begin{align}
\Im G^{R}_{\eta\xi,0}(E;0,0) \rightarrow \Im G^{R}_{\eta\xi,0}(E;0,0)- \Im G^{R}_{\eta\xi,0}(E_0;0,0)=\frac{m^2}{4\pi}\alpha \left[\log\left(\frac{v}{v_0}\right)+\Re\Psi\left(-i\frac{\alpha}{v}\right)-\Re\Psi\left(-i\frac{\alpha}{v_0}\right)\right].
\end{align}
}

Having a finite $G^R_{\eta\xi,0}(E;0,0)$, we turn to bound-state contributions in \eqref{eq:selfconsistentretarded}. Focusing on energies close to their binding, the retarded solutions are given by:
\begin{align}
G^R_{\eta \xi,0}(E\sim E_n,0,0) = i \frac{ |\psi_{n}(0)|^2}{E-E_n+ i \epsilon} + \text{regular}\;, \quad G^R_{\eta \xi}(E\sim E_n;0,0) \simeq i \frac{|\psi_{n}(0)|^2}{(E-E_n)+ ic|\psi_{n}(0)|^2} \;.\label{eq:reaterdedbounddressed}
\end{align}
Let us define the decay rate of a $s$-wave bound state: $\Gamma^{\text{dec}}_n\equiv 2c|\psi_{n}(0)|^2$. In terms of this, the spectral function computed from Eq.~\eqref{eq:reaterdedbounddressed} at energies close to the binding is:
\begin{align}
G^{\rho}_{\eta \xi}(E\sim E_n;0,0)\simeq |\psi_{n}(0)|^2 \frac{\Gamma_n^\text{dec}}{(E-E_n)^2 + (\Gamma_n^\text{dec}/2)^2} \rightarrow \frac{2}{c} \quad \text{, for } E=E_n. \label{eq:bBW}
\end{align}
In the limit $\Gamma_n^\text{dec} \rightarrow 0$, the Breit-Wigner function correctly reduces to the on-shell delta function $2\pi \delta(E-E_n)$ in \eqref{eq:Grhoetaxi0}. Exactly on the pole location, we observe a universal height of all $s$-wave bound states in the spectral function $G^{\rho}_{\eta \xi}(E = E_n;0,0)= \frac{2}{c}$. Let us turn to the scattering states, i.e. $E>0$, and let us assume that the (Yukawa) potential features a single bound state close to the threshold $E_n \sim 0$. Neglecting the imaginary part of $G^R_{\eta \xi,0}(E;0,0)$ compared to the resonantly enhanced real part, the spectral function can be approximated as:
\begin{align}
G^{\rho}_{\eta \xi}(E \gtrsim 0;0,0) \simeq \frac{m^2}{2\pi}\sqrt{\frac{E}{m}} \frac{S(v)}{1+\frac{c S(v)}{2(\sigma v)_\text{uni}}} \rightarrow \frac{2}{c} \quad \text{, for } \frac{c S(v)}{2(\sigma v)_\text{uni}} \gg 1 \;.\label{eq:specpositive}
\end{align}
Suppose that we are close to a zero energy resonance where $cS$ is much larger than the s-wave unitarity cross section $(\sigma v)_\text{uni}=4\pi/(m^2v)$. In this case, the spectral function approaches a constant for small relative velocity.
This matches the maximum height of the bound state close to the threshold. Thus, the spectral function allows for an intuitive interpretation of the unitarization: due to the presence of a width induced by annihilation, the maximum height of a bound state at zero energy is $2/c$; the same value the spectral function approaches for positive energies in the zero velocity limit in the on-resonance case.

\begin{claim}
However, a naive replacement $G^\rho_{\eta\xi,0} \rightarrow G^\rho_{\eta\xi}$ in the collision term~\eqref{eq:spectral}, which relies on the KMS relation, is \emph{incorrect} due to the violation of discussed symmetries. It would lead to two inconsistencies:
\begin{enumerate}
    \item Since bound states give rise to Breit-Wigner-type spectral shapes, see Eq.~\eqref{eq:bBW}, the energy $E$ integral appearing in the collision term~\eqref{eq:spectral} becomes divergent at large negative energies. In this regime, the exponential growth of the thermal distribution function overwhelms the Breit-Wigner suppression. Consequently, in the presence of bound states, the collision term diverges if one replaces $G^\rho_{\eta\xi,0}$ by the self-consistently computed spectral function $G^\rho_{\eta\xi}$.
    \item The resulting expression for the unitarized Sommerfeld enhancement, corresponding to Eq.~\eqref{eq:specpositive}, does not reproduce the result of Ref.~\cite{Blum:2016nrz} when the same approximations are applied (retaining only annihilation effects and the real part of the Green's function). In particular, Ref.~\cite{Blum:2016nrz} exhibits a squared dependence on the denominator of the corresponding term (from squaring the wave function), in contrast to the linear dependence obtained here.
\end{enumerate}
\end{claim}
In summary, when invoking the KMS condition---which strictly excludes annihilation contributions from the computation of the four-point function~\eqref{eq:defgppmmfree} due to the violation of global $U(1)$ symmetries---the substitution $G^\rho_{\eta\xi,0} \rightarrow G^\rho_{\eta\xi}$ in Eq.~\eqref{eq:spectral} is not justified. An additional insight from this analysis is that, although the Sommerfeld enhancement derived from the spectral function coincides with that obtained from the Schr\"odinger equation for real-valued potentials, the two approaches generally yield different results in the presence of complex potentials\footnote{The reason for this is that the spectral function is proportional to the total cross section, that includes also elastic scattering induced by the complex delta potential. Elastic scattering should not enter into the collision term that changes number. A direct way to see this is to add the unitarized elastic and inelastic cross sections in \cite{Flores:2024sfy} that leads to the same answer when taking the imaginary part of the Green's function. One could therefore subtract the elastic piece in \eqref{eq:specpositive} for the scattering states, giving the correct answer as in \cite{Blum:2016nrz}, however, how to perform such ad-hoc subtraction for the bound states is unclear.} .

Therefore, we see the only possible way to obtain the correct self-consistent solution by directly solving for $G_{\eta \xi}^{++--}$ without relying on the KMS relation, to which we turn next. And in fact, we will show that when doing such self-consistent computation, neither the bound states acquire a Breit-Wigner energy shape in the collision term (1.), nor the scattering state solution deviates by a square in the denominator of the regulated Sommerfeld enhancement from \cite{Blum:2016nrz} (2.).

\section{Self-consistent four-point functions from pNREFT}
\label{sec:pNREFTapproach}

A straightforward way to compute $G^{++--}_{\eta \xi}$ self-consistently is to derive and solve the equations of motion for the coupled four-point correlation functions from the NREFT action on the CTP contour in Eq.~\eqref{eq:NRaction}, including annihilation. The four-point function is coupled to the six-point function, reflecting an infinite hierarchy of correlators. By truncating this hierarchy at the four-point function level, one obtains a closed set of equations that can be solved analytically. This first approach, based on a direct computation in NREFT, is presented in Section~\ref{sec:NRpaircreation} and Appendix~\ref{app:NREFT}.

In the second approach, we equate the four-point correlation function $G_{\eta \xi}$ in NREFT to the corresponding four-point correlation function in \emph{potential non-relativistic effective field theory} (pNREFT). As the second approach based on pNREFT is computationally simpler, we shall present it in this section. Note that both our results, obtained from NREFT and pNREFT, are in agreement.

\subsection{pNREFT on CTP contour}
By projecting the NREFT action on the CTP contour in Eq.~\eqref{eq:NRaction} into the particle-antiparticle sub-space, we get for the pNREFT action on the CTP contour:
\begin{align}
S_{\text{pNR}} =\int\limits_{x^0\in \mathcal{C}} \text{d}^4x \text{d}^3r \; S^\dagger(x,\mathbf{r}) [ i \partial_{x^0} - h ] S(x,\mathbf{r}) + i \int\limits_{x^0, y^0 \in \mathcal{C}} \text{d}^4x  \text{d}^4y \text{d}^3r \; S^\dagger(x,\mathbf{r}) \delta(\mathbf{r}) \Gamma(x,y) S(y,\mathbf{r}) \;,
\end{align}
where $S$ is the two-body field, now $\mathbf{x}$, $\mathbf{y}$ denote center-of-mass (CM) coordinates and $\mathbf{r}$ is the relative coordinate, and $\Gamma(x,y)$ is the same as in Eq.~\eqref{eq:annihop}, which accounts for annihilation of the pairs. The resulting EoMs for the two-body field correlator,
\begin{align}
G(x,y;\mathbf{r},\mathbf{r}^\prime) \equiv \langle T_\mathcal{C} S(x,\mathbf{r}) S^{\dagger}(y,\mathbf{r}^\prime) \rangle \;, \label{eq:defpNRcorr}
\end{align}
on the CTP contour are closed and given by:
\begin{align}
(i \partial_{x^0} - h_x)G(x,y;\mathbf{r},\mathbf{r}^\prime) &= i \delta^{4}_\mathcal{C}(x-y) \delta(\mathbf{r} -\mathbf{r}^\prime) - i \int\limits_{z^0 \in \mathcal{C}} \text{d}^4 z \delta(\mathbf{r}) \Gamma(x,z) G(z,y;\mathbf{r},\mathbf{r}^\prime) \;, \label{eq:firsteom}\\
(-i \partial_{y^0} - h_y^\prime)G(x,y;\mathbf{r},\mathbf{r}^\prime) &= i \delta^{4}_\mathcal{C}(x-y) \delta(\mathbf{r} -\mathbf{r}^\prime) - i \int\limits_{z^0 \in \mathcal{C}} \text{d}^4 z G(x,z;\mathbf{r},\mathbf{r}^\prime) \Gamma(z,y)  \delta(\mathbf{r}^\prime)\;, \label{eq:secondeom}
\end{align}
where 
\begin{align}
h_x= - \frac{\Delta_\mathbf{r}}{m}-\frac{\Delta_\mathbf{x}}{4m}+V(r), \quad h_y^\prime= - \frac{\Delta_{\mathbf{r}^\prime}}{m}-\frac{\Delta_\mathbf{y}}{4m}+V(r^\prime) \;,
\end{align}
and $V(r)$ denotes the static potential. The four-point correlator in NREFT, $G_{\eta \xi}^{++--}$, can be estimated from the four-point-correlator in pNREFT, $G^{+-}$, as (see, similarly Ref.~\cite{Brambilla:2017zei}):
\begin{align}
G_{\eta \xi}^{++--}(t, \mathbf{x}_1, \mathbf{x}_2;t^\prime,\mathbf{x}_3,\mathbf{x}_4) \approx G^{+-}(x,y;\mathbf{r}, \mathbf{r}^\prime) \;.
\end{align}
Effectively, we replaced the bilinear in NREFT at equal times, $\eta(t,\mathbf{x}_1)\xi^\dagger(t,\mathbf{x}_2)$, by the interpolating field $S(x,\mathbf{r})$ in CM and relative coordinates. This reduces the problem to solving $G^{+-}$ self-consistently. For this purpose, we introduce the Green's function $G_0$ without annihilation:
\begin{align}
(i \partial_{x^0} - h_x)G_0(x,y;\mathbf{r},\mathbf{r}^\prime) &= i \delta^{4}_\mathcal{C}(x-y) \delta(\mathbf{r} -\mathbf{r}^\prime) \;, \label{eq:freepnr1} \\
(-i \partial_{y^0} - h_y^\prime)G_0(x,y;\mathbf{r},\mathbf{r}^\prime) &= i \delta^{4}_\mathcal{C}(x-y) \delta(\mathbf{r} -\mathbf{r}^\prime) \;. \label{eq:freepnr2}
\end{align}
In Appendix~\ref{app:freepNR} we provide the solution to these equations. Knowing $G_0$, the full solution of \eqref{eq:firsteom} and \eqref{eq:secondeom} including annihilation is formally given by:
\begin{align}
G(x,y;\mathbf{r},\mathbf{r}^\prime)= G_0(x,y;\mathbf{r},\mathbf{r}^\prime)- \int\limits_{z^0,w^0 \in \mathcal{C}} \text{d}^4z \text{d}^4w \text{d}^3\bar{\mathbf{r}} \; G_0(x,z;\mathbf{r}, \bar{\mathbf{r}}) \delta(\bar{\mathbf{r}}) \Gamma(z,w) G(w,y;\bar{\mathbf{r}}, \mathbf{r}^\prime) \label{eq:formalsol}\;.
\end{align}
We solve this equation explicitly in the following in Section \ref{sec:integral}, given $G_0$. For an interacting initial value problem we solve the pNREFT equations \eqref{eq:firsteom} and \eqref{eq:secondeom} analytically in Section \ref{sec:exact}.

For both cases, we work with Wigner coordinates for the \emph{Center-of-Mass} variables as:
\begin{align}
\mathbf{X}\equiv(\mathbf{x}+\mathbf{y})/2 \;, \quad  \mathbf{R} \equiv \mathbf{x}-\mathbf{y} \;, \quad T \equiv (x^0+y^0)/2 \;, \quad t \equiv x^0-y^0\;.
\end{align}
We define the correlator dependence in terms of these as:
\begin{align}
G(T,t,\mathbf{X},\mathbf{R};\mathbf{r},\mathbf{r}^\prime) \equiv G\left(T+\frac{t}{2},\mathbf{X}+\frac{\mathbf{R}}{2},T-\frac{t}{2},\mathbf{X}-\frac{\mathbf{R}}{2};\mathbf{r},\mathbf{r}^\prime \right) = G(x,y;\mathbf{r},\mathbf{r}^\prime) \;.
\end{align}
Throughout this work, it is assumed that $G$ does not depend on $\mathbf{X}$ in a spatially homogeneous and isotropic system and the correlator dependence on it is dropped. The Fourier-transformed correlator is denoted by:
\begin{align}
G(T,P;\mathbf{r},\mathbf{r}^\prime) \equiv G(T,\omega,\mathbf{P}; \mathbf{r},\mathbf{r}^\prime) = \int \text{d}t \text{d}^3 R \; e^{i(\omega t- \mathbf{P}\cdot \mathbf{R})} G(T,t,\mathbf{R};\mathbf{r},\mathbf{r}^\prime)  \;, \label{eq:transformed}
\end{align}
where we frequently use the notation $P=(P^0,\mathbf{P}) = (\omega,\mathbf{P})$.

\subsection{Analytic solution based on integral form}
\label{sec:integral}

Applying the Wigner and Fourier transformation in Eq.~\eqref{eq:transformed} to Eq.~\eqref{eq:formalsol}, we get, at the leading order in the gradient expansion, a coupled set of correlators, given by:
\begin{align}
G^{+-}(P;\mathbf{r},\mathbf{r}^\prime) &= G^{+-}_0(P;\mathbf{r},\mathbf{r}^\prime) - c \big[ G^{R}_0(P;\mathbf{r},0) G^{+-}(P;0,\mathbf{r}^\prime) - G^{+-}_0(P;\mathbf{r},0) G^{A}(P;0,\mathbf{r}^\prime)\big] \label{eq:pm}\;,\\
G^{A}(P;\mathbf{r},\mathbf{r}^\prime) &= G^{A}_0(P;\mathbf{r},\mathbf{r}^\prime) + c G_0^A(P;\mathbf{r},0) G^{A}(P;0,\mathbf{r}^\prime)\;.\label{eq:adv}
\end{align} 
 This is a system of linear equations. The solution for our target components at the origin reads:
\begin{align}
G^{+-}(P;0,0) &= \frac{1}{\big[1+c G^R_0(P;0,0)\big]\big[1-c G^A_0(P;0,0)\big]} G_0^{+-}(P;0,0)\;, \\
G^{A}(P;0,0) &= \frac{1}{1-c G^A_0(P;0,0)} G_0^{A}(P;0,0) \;. \label{eq:resummedadvanced}
\end{align}
The advanced correlator results in the same spectral function as discussed earlier in the NREFT approach. However, the actual quantity which enters the annihilation collision term \eqref{eq:number} is $G^{+-}$. With the basic relation $G^A_0 =- (G_0^R)^\star$ we arrive at the final form of one of our main results:
\begin{claim}
\begin{align}
 G^{+-}(P;0,0) &= \frac{1}{\big|1+c G^R_0(P;0,0)\big|^2} \times G_0^{+-}(P;0,0) \;.  \label{eq:res} 
\end{align}
\end{claim}
In Appendix~\ref{app:freepNR}, we show that for a stationary state, in the dilute regime and under the assumption of kinetic equilibrium, the statistical correlator for scattering and bound states can be written as:
\begin{align}
G_0^{+-}(P;0,0)=  e^{-\beta(2m + \mathbf{P}^2/(4m) + E )} \times \begin{cases}
			(n_s/n_s^\text{eq})^2 \frac{m^2}{2\pi} \sqrt{\frac{E}{m}} S(v) , & \text{for } E > 0 \;, \\
            (n_n/n_n^\text{eq}) (2\pi)\delta(E-E_n) |\psi_n(0)|^2 , & \text{for } E \sim E_n \;.
		 \end{cases} \label{eq:cases}
\end{align}
Inserting the scattering state part, $E>0$, together with \eqref{eq:res} into the NR EoM in Eq.~\eqref{eq:number}, we obtain the self-consistently computed Sommerfeld enhancement compatible with \cite{Blum:2016nrz}, which respects the s-wave partial-wave unitarity bound. The derivation in Ref.~\cite{Blum:2016nrz} is based on the stationary solution of the Schrödinger equation for scattering states in vacuum, while here we confirm the insertion into the Boltzmann equation directly from thermal field theory. Notice the appearance of a squared denominator in \eqref{eq:res} in contrast to the spectral function approach in Section~\ref{sec:KMS}.

For bound states, the retarded propagator at the origin and $E\sim E_n$ can be written as:
\begin{align}
G^R_0(E\sim E_n,0,0) = i \frac{ |\psi_{n}(0)|^2}{E-E_n+ i \epsilon} + \text{regular} \;.
\end{align}
Inserting this into Eq.~\eqref{eq:res}, one observes that bound states drop out of the collision term:
\begin{align}
\frac{1}{\big|1+c G^R_0(E\sim E_n;0,0)\big|^2} \times \delta(E-E_n) = 0 \;.
\end{align}
This is due to the fact that $G_0^{+-}\propto \delta(E-E_n)$ for bound states in \eqref{eq:cases}. At first sight, this result may appear counterintuitive: once short-distance annihilation is included beyond leading order, bound-state contributions drop out of the Boltzmann equation. This apparent puzzle can be understood as follows. Up to this point, we have neglected gradients (the ``diamond operator") in Eq.~\eqref{eq:pm} and \eqref{eq:adv}, implying that we restricted ourselves to a \emph{stationary solution} in Eq.~\eqref{eq:res}. 

This is because the LO gradient expansion assumes that all correlators depend on $x-y$ only (as in vacuum), implying the time translation invariance. While this approximation is consistent for scattering states, it is insufficient for bound states, which are intrinsically unstable and therefore necessarily exhibit time-dependent behavior due to their decay. 

To properly capture the dynamics of bound states, it is therefore essential to retain time gradients. By fully including those gradients, the more general version of Eq.~\eqref{eq:pm} reads:
\begin{align}
&G^{+-}(T,\omega,\mathbf{P};0,0) = G^{+-}_0(T,\omega,\mathbf{P};0,0) - \\ & c \big[ G_0^R(T,\omega,\mathbf{P};0,0) e^{-i\diamond  } G^{+-}(T,\omega,\mathbf{P};0,0) - G_0^{+-}(T,\omega,\mathbf{P};0,0) e^{-i\diamond   }  G^A(T,\omega,\mathbf{P};0,0)\big] \;. \nonumber
\end{align}
The ``diamond operator", which incorporates the gradients, is given by:
\begin{equation}
    \diamond = \frac{1}{2} \left(\partial_T^{(l)} \partial_\omega^{(r)} - \partial_\omega^{(l)} \partial_T^{(r)} \right),
\end{equation}
where the superscript $l$ ($r$) labels that the derivative is acting on the correlator to the left (right). As always, we neglect spatial $\mathbf{X}$ dependence. In such a case, it is shown in Appendix~\ref{app:freepNR} that the retarded and advanced correlation functions are Wigner time $T$ independent. Further, we assume that the initial density matrix is diagonal in momentum space, implying $T$ independence of $G_0^{+-}$ (see Appendix~\ref{app:freepNR}). Collectively, these considerations yield the considerably simplified equation:
\begin{align}
&G^{+-}(T,\omega,\mathbf{P};0,0) = G^{+-}_0(\omega,\mathbf{P};0,0) - \label{eq:gpmpnreft}\\ & c \big[ G_0^R(E;0,0) e^{\frac{i}{2}\partial_E^{(l)} \partial_T^{(r)} } G^{+-}(T,\omega,\mathbf{P};0,0) - G_0^{+-}(\omega,\mathbf{P};0,0) G^A(E;0,0)\big] \;. \nonumber
\end{align}
We made use of the fact that the retarded and advanced correlators depend on a specific combination of $E\equiv \omega - \mathbf{P}^2/(4m)$, as it was also the case in Section \ref{sec:KMS} for the NREFT version.

We start with bound states, where the statistical correlator, in the diagonal density matrix limit, is given by:
\begin{align}
G_{0}^{+-}(\omega, \mathbf{P};0, 0)= \sum_n f_n(2m + \omega) (2\pi) \delta(\omega - \mathbf{P}^2/(4m) - E_n) |\psi_n(0) |^2 \;, \text{ for } E<0.
\end{align}
This expression is more general than Eq.~\eqref{eq:cases} in kinetic equilibrium. The phase-space distribution $f_n$ of a s-wave bound state with principle quantum number $n$ is normalized to its number density $n_n$ as:
\begin{align}
n_n= \int \frac{\text{d}^4P}{(2\pi)^4} f_n(2m + P^0) (2\pi) \delta(P^0 - \mathbf{P}^2/(4m) - E_n)\;.
\end{align}
For bound states, the last term in the brackets in Eq.~\eqref{eq:gpmpnreft} cancels the first term
\begin{align}
 G^{+-}_0(\omega,\mathbf{P};0,0)  +   c G_0^{+-}(\omega,\mathbf{P};0,0)  G^A(E;0,0) =0\;. 
\end{align}
This cancellation appears only for bound states, where $G_0^{+-}\propto \delta(E-E_n)$ and the self-consistently computed advanced correlator is given in \eqref{eq:resummedadvanced}. Consequently, for bound states only the first gradient term in Eq.~\eqref{eq:gpmpnreft} contributes and can be computed to all orders, giving for $E\sim E_n$:
\begin{align} 
G^{+-}(T,\omega,\mathbf{P};0,0)  &= - i \underbrace{c |\psi_{n}(0)|^2}_{ = \Gamma^\text{dec}_n/2} \sum_{m=0}^{\infty} \frac{i^m}{m! 2^m} \left(\partial_E^m \frac{1}{E-E_n + i \epsilon} \right) \partial_T^m G^{+-}(T,\omega,\mathbf{P};0,0) \nonumber \\
&=-i \frac{\Gamma^\text{dec}_n}{2} [E- E_n + i \partial_T/2]^{-1} G^{+-}(T,\omega,\mathbf{P};0,0)\;.  \label{eq:boundgpmgrad}
\end{align}
The energy gradient converted the exponential series into a geometric series. Notice that $E\sim E_n$ implies here a linearization to get to the leading order bound state decay width $\Gamma^\text{dec}_n$, which ignores other contributions from suppressed other bound and scattering state contributions. Formally inverting, and requiring $G^{+-}(T,\omega,\mathbf{P};0,0) $ to be real, leads to two conditions, given for $E \sim E_n$ by:
\begin{align}
\partial_T G^{+-}(T,\omega,\mathbf{P};0,0)  &= - \Gamma_n^\text{dec} G^{+-}(T,\omega,\mathbf{P};0,0) \;, \\
(E-E_n) G^{+-}(T,\omega,\mathbf{P};0,0)  &= 0 \;.
\end{align}
The first equation leads to exponential decay in Wigner time $T$, while the second equation implies that the solution must be on-shell. A solution, which is compatible with both requirements is:
\begin{claim}
\begin{align}
G^{+-}(T,P;0,0) &= \sum_n e^{-\Gamma_n^\text{dec} T}  f_n(T_0,2m + \omega) (2\pi) \delta(\omega - \mathbf{P}^2/(4m) - E_n) |\psi_n(0) |^2 \;, \text{ for } E<0. \label{eq:solnegative}
\end{align}
\end{claim}
To be consistent with the differential equations, $\psi_n$ must be an Eigenvector of $h$ plus the imaginary annihilation contribution, with complex energy eigenvalue given by real $E_n$ minus $i \Gamma_n^\text{dec}$, and properly normalized. In this way, the wavefunction at the origin also acquires corrections from the decay. Notice that $\Gamma_n^\text{dec}$ is only the leading decay width, and originates from the expansion $E\sim E_n$ in Eq.~\eqref{eq:boundgpmgrad}. In Section~\ref{sec:exact}, we give a complete definition. 

The above solution is noteworthy. One might have naively expected from the spectral function that bound states acquire a finite decay width, resulting in a Breit–Wigner energy profile. Remarkably, this does not occur and all s-wave bound states remain on-shell in their decay out-of-equilibrium, due to $\delta(\omega - \mathbf{P}^2/(4m) - E_n) = \delta(E-E_n)$.

A closely related phenomenon has been observed for an elementary particle undergoing decay when the Kadanoff–Baym equations are solved exactly~\cite{Anisimov:2008dz, Garbrecht:2011xw}. In such cases, the decay width also does not enter the dispersive part of the out-of-equilibrium contribution: the excitation remains strictly on shell.
As we have not yet included pair creation, our system considered in this Section never reaches chemical equilibrium. As we shall see later, when including pair creation, the solution for bound states at long times will be the equilibrium solution where a Breit-Wigner shape appears, consistent with the fluctuation–dissipation theorem (see Section~\ref{sec:exact}). However, in equilibrium, the collision term including pair creation (see Section~\ref{sec:NRpaircreation}) vanishes, and therefore does not cause integration problems over Breit-Wigner energy shapes.

For comparison, let us insert the negative energy solution in Eq.~\eqref{eq:solnegative} and the stationary solution for positive energies in Eq.~\eqref{eq:res} into the particle number density Eq.~\eqref{eq:number}, arriving at:
\begin{align}
\dot{n}_\eta = - \langle \sigma v \rangle n^2_s - \sum_n \Gamma_n^\text{dec}  e^{-\Gamma_n^\text{dec} T} n_n(T_0) \;,\label{eq:Boltzmannresult}
\end{align}
where $\langle \sigma v \rangle$ denotes the thermal average of the self-consistently computed Sommerfeld-enhanced annihilation cross section and we neglected finite width corrections to the bound state wave functions. This equation is consistent with expectations from independent Boltzmann equations for scattering states and bound states which can annihilate only:
\begin{align}
\dot{n}_s = - \langle \sigma v \rangle n^2_s\;, \quad \dot{n}_n= -\Gamma_n n_n \;.
\end{align} 
First solving all bound state equations for $n_n(T)$, inserting this back into the bound state differential equations, summing over all, and defining the total number as $n_\eta = n_s + \sum_n n_n$, gives \eqref{eq:Boltzmannresult}. Within the Boltzmann equation framework, the number density of bound states must be postulated in this form. By contrast, a self-consistent treatment of bound states in thermal field theory reproduces the same Boltzmann equation that is conventionally employed, when restricting to the leading order contribution of the decay width. 

A crucial assumption to arrive at this semi-classical Boltzmann equation description was to assume that $G_0^{+-}$ does not depend on Wigner time $T$, which is only the case for a diagonal density matrix in momentum space, or under the rotating wave approximation valid for long times after the initial time (see Appendix~\ref{app:freepNR}). In this way, we neglect the coherence effects for scattering and bound states. We turn now to an analytic solution for essentially arbitrary initial conditions.

\subsection{Analytic solution including pair creation for initial value problem}
\label{sec:exact}
Sofar we solved the pNREFT equations in integral form, based on the gradient method, which requires the knowledge of $G_0$. Here, we turn back to the pNREFT equations \eqref{eq:firsteom} and \eqref{eq:secondeom} in differential form. These will be analytically solved, now instead for an initial value problem. By obtaining an exact analytic solution in this Section, we will reconfirm our previous gradient-method-based solutions under consistent limits. From now on, we will work with a general and complex
\begin{align}
\Gamma(x,y) =
\begin{pmatrix}
\Gamma^{++}(x-y) & \Gamma^{+-}(x-y) \\
\Gamma^{-+}(x-y) & \Gamma^{--}(x-y)
\end{pmatrix} \;, \label{eq:generalmatching}
\end{align}
which includes pair creation effects. The only assumption on this quantity is that the large thermal environment is in equilibrium, which implies translational invariance of $\Gamma=\Gamma(x-y)$.

\subsubsection{Derivation}
\label{sec:pnrderivation}
To obtain an analytic solution to the pNREFT equations \eqref{eq:firsteom} and \eqref{eq:secondeom}, we start by following Ref.~\cite{Berges:2004yj} to express the full correlator on the CTP contour,
\begin{align}
G(x,y;\mathbf{r},\mathbf{r}^\prime)&= G^{-+}(x,y;\mathbf{r},\mathbf{r}^\prime) \theta_\mathcal{C}(x^0-y^0) +G^{+-}(x,y;\mathbf{r},\mathbf{r}^\prime) \theta_\mathcal{C}(y^0-x^0) \nonumber \\
&=G^F(x,y;\mathbf{r},\mathbf{r}^\prime) + \frac{1}{2} \sign_\mathcal{C}(x^0-y^0) G^\rho(x,y;\mathbf{r},\mathbf{r}^\prime)\;,
\end{align}
in terms of the already introduced spectral function $G^\rho$ and the anti-commutator correlator $G^F$, which in terms of the two-body field is defined as:
\begin{align}
G^F(x,y;\mathbf{r},\mathbf{r}^\prime) = \frac{1}{2} \langle \{S(x,\mathbf{r}),S^\dagger(y,\mathbf{r}^\prime) \}\rangle \;.
\end{align}
The advantage of working with $G^F$ and $G^\rho$ instead of $G^{+-}$ and the retarded and the advanced component as before, is that initial conditions become manifest in the solution, clear from below.

Expressing the pNREFT equations \eqref{eq:firsteom} and \eqref{eq:secondeom} in terms of $G^F$ and $G^\rho$, results (without approximations) in:
\begin{align}
\left( i \partial_{x_0} - h_x\right) G^F(x,y;\mathbf{r},\mathbf{r}^\prime) &= -i \int_{t_0}^{x^0} \text{d}z^0 \int \text{d}^3 z \delta(\mathbf{r}) \Gamma^\rho(x-z) G^F(z,y;\mathbf{r},\mathbf{r}^\prime) \nonumber \\ &+ i\int_{t_0}^{y^0} \text{d}z^0 \int \text{d}^3 z \delta(\mathbf{r}) \Gamma^F(x-z) G^\rho(z,y;\mathbf{r},\mathbf{r}^\prime) \;, \label{eq:Gfdef}\\
\left( i \partial_{x_0} - h_x \right) G^\rho(x,y;\mathbf{r},\mathbf{r}^\prime) &= -i \int_{y^0}^{x^0} \text{d}z^0 \int \text{d}^3 z  \delta (\mathbf{r}) \Gamma^\rho(x-z) G^\rho(z,y;\mathbf{r},\mathbf{r}^\prime) \;, \label{eq:specdef}
\end{align}
for initial time $t_0$. For solving these equations analytically, we make one assumption: $G^F$ and $G^\rho$ are assumed to be translationally invariant in the spatial center-of-mass coordinates, i.e., $G^F$ and $G^\rho$ do not depend on the Wigner coordinate $\mathbf{X}=(\mathbf{x}+\mathbf{y})/2$. Under this assumption, we follow closely the method in Ref.~\cite{Anisimov:2008dz} which lead to analytic solutions for the Kadanoff-Baym two-point function equations, and adapt their method to obtain analytic solutions for pNREFT four-point functions.

The spectral function satisfies the boundary condition:
\begin{align}
G^{\rho}(x^0,x^0,\mathbf{P},\mathbf{r},\mathbf{r}^\prime)=\delta(\mathbf{r}-\mathbf{r}^\prime)\;,\label{eq:boundarygrho}
\end{align}
which follows from the canonical equal-time commutation relation of the two-body fields:
\begin{align}
[S(x^0,\mathbf{x},\mathbf{r}),S^\dagger(x^0,\mathbf{y},\mathbf{r}^\prime)]=\delta(\mathbf{x}-\mathbf{y})\delta(\mathbf{r}-\mathbf{r}^\prime)\;.
\end{align}
Using the boundary condition, we show in Appendix~\ref{app:freepNR} that $G^\rho$ is time translation invariant (does not depend on the Wigner-time $T=(x^0+y^0)/2$). Therefore, Eq.~\eqref{eq:specdef} simplifies to:
\begin{align}
\left( i \partial_t - \frac{\mathbf{P}^2}{4m} + \frac{\Delta_\mathbf{r}}{m} - V(r)\right) G^\rho(t,\mathbf{P},\mathbf{r},\mathbf{r}^\prime) =-i \int_0^t \text{d}t^\prime \delta(\mathbf{r}) \Gamma^\rho(t-t^\prime,\mathbf{P}) G^\rho(t^\prime,\mathbf{P};\mathbf{r},\mathbf{r}^\prime)\;. \label{eq:Grhofull}
\end{align}
with time difference variable $t=x^0-y^0$. This is a linear Volterra integro-differential equation with convolution kernel. The solution can be obtained by applying the Laplace transformation to this time variable, where the time derivative term transforms as
\begin{align}
\int_0^\infty\text{d}t e^{-st} i\partial_t G^\rho(t,\mathbf{P},\mathbf{r},\mathbf{r}^\prime) = -i\delta(\mathbf{r}-\mathbf{r}^\prime) + s i \tilde{G}^\rho(s,\mathbf{P},\mathbf{r},\mathbf{r}^\prime) \;. 
\end{align}
We used partial integration and the boundary condition in \eqref{eq:boundarygrho}, $G^\rho(t=0,\mathbf{P},\mathbf{r},\mathbf{r}^\prime)=\delta(\mathbf{r}-\mathbf{r}^\prime)$. Laplace transforming \eqref{eq:Grhofull}, gives:
\begin{align}
\left( is- \frac{\mathbf{P}^2}{4m} + \frac{\Delta_\mathbf{r}}{m} - V(r) +i \delta(\mathbf{r}) \tilde{\Gamma}^\rho(s,\mathbf{P}) \right)\tilde{G}^\rho(s,\mathbf{P},\mathbf{r},\mathbf{r}^\prime)= i \delta(\mathbf{r}-\mathbf{r}^\prime) \;.
\end{align}
Note that the spectral function in Laplace space is a Green's function, and therefore the self-consistent solution is given by:
\begin{align}
\tilde{G}^\rho(s,\mathbf{P},\mathbf{r},\mathbf{r}^\prime)&= \tilde{G}^\rho_0(s,\mathbf{P},\mathbf{r},\mathbf{r}^\prime)-\tilde{\Gamma}^\rho(s,\mathbf{P}) \tilde{G}^\rho_0(s,\mathbf{P},\mathbf{r},0)\tilde{G}^\rho(s,\mathbf{P},0,\mathbf{r}^\prime) \;, \\
\tilde{G}^\rho(s,\mathbf{P},0,\mathbf{r}^\prime) &= \frac{\tilde{G}^\rho_0(s,\mathbf{P},0,\mathbf{r}^\prime)}{1+ \tilde{\Gamma}^\rho(s,\mathbf{P})\tilde{G}^\rho_0(s,\mathbf{P},0,0)} \;.
\end{align}
We turn to the solution of Eq.~\eqref{eq:Gfdef}. We start with the homogeneous part, denoted by $G_h^F$, given by:
\begin{align}
\left( i \partial_{x^0} - \frac{\mathbf{P}^2}{4m} + \frac{\Delta_\mathbf{r}}{m} - V(r)\right) G^F_h(x^0,y^0,\mathbf{P};\mathbf{r},\mathbf{r}^\prime) = -i \int_{0}^{x^0} \text{d}z^0 \delta(\mathbf{r}) \Gamma^\rho(x^0-z^0) G^F_h(z^0,y^0,\mathbf{P};\mathbf{r},\mathbf{r}^\prime) \;,
\end{align}
where from now on we set the initial time to $t_0=0$. One can notice that this homogeneous equation has a similar structure as Eq.~\eqref{eq:Grhofull}, but now $y^0$ playing the role of a parameter. Laplace transforming with respect to $x^0$ gives:
\begin{align}
    \left( is- \frac{\mathbf{P}^2}{4m} + \frac{\Delta_\mathbf{r}}{m} - V(r)+i \delta(\mathbf{r}) \tilde{\Gamma}^\rho(s,\mathbf{P}) \right)\tilde{G}^F_h(s,y^0,\mathbf{P};\mathbf{r},\mathbf{r}^\prime) = i \tilde{G}^F_h(x^0=0,y^0,\mathbf{P};\mathbf{r},\mathbf{r}^\prime) \;.
\end{align}
Recalling that the spectral function is the Green's function of the operator to the left in Laplace-space, the solution can be expressed as:
\begin{align}
\tilde{G}^F_h(s,y^0,\mathbf{P};\mathbf{r},\mathbf{r}^\prime) = \int \text{d}^3r_1 \tilde{G}^\rho(s,\mathbf{P};\mathbf{r},\mathbf{r}_1) \tilde{G}^F_h(x^0=0,y^0,\mathbf{P};\mathbf{r}_1,\mathbf{r}^\prime) \;.
\end{align}
After applying inverse Laplace-transformation we get for the homogeneous part in coordinate space:
\begin{align}
G^F_h(x^0,y^0,\mathbf{P};\mathbf{r},\mathbf{r}^\prime) = \int \text{d}^3r_1 G^\rho(x^0,\mathbf{P};\mathbf{r},\mathbf{r}_1) G^F_h(x^0=0,y^0,\mathbf{P};\mathbf{r}_1,\mathbf{r}^\prime) \;. \label{eq:hom1}
\end{align}
Note that the relative time argument of the spectral function is now evaluated at $x^0$. To eliminate the parametric dependence of the initial condition on $y^0$, we use, from the basic definition of the correlator, that $G^{F}(x,y;\mathbf{r},\mathbf{r}^\prime)= [G^{F}(y,x;\mathbf{r}^\prime,\mathbf{r})]^\dagger $, implying:
\begin{align}
\int \text{d}^3r_1 G^\rho(x^0,\mathbf{P};\mathbf{r},\mathbf{r}_1) G^F_h(x^0=0,y^0,\mathbf{P};\mathbf{r}_1,\mathbf{r}^\prime) = \int \text{d}^3r_2 [G^F_h(y^0=0,x^0,\mathbf{P};\mathbf{r}_2,\mathbf{r}]^\dagger [G^\rho(y^0,\mathbf{P};\mathbf{r}^\prime,\mathbf{r}_2)]^\dagger \;. 
\end{align}
Setting $x^0=0$, and using the boundary condition of the spectral function in \eqref{eq:boundarygrho}, gives the following relation:
\begin{align}
 G^F_h(x^0=0,y^0,\mathbf{P};\mathbf{r},\mathbf{r}^\prime)  = \int \text{d}r_2 [G^F_h(y^0=0,x^0=0,\mathbf{P};\mathbf{r}_2,\mathbf{r}]^\dagger [G^\rho(y^0,\mathbf{P};\mathbf{r}^\prime,\mathbf{r}_2)]^\dagger \;.
\end{align}
Inserting this back into Eq.~\eqref{eq:hom1}, shows that the time evolution of the homogeneous part is entirely set by a single initial condition at $x^0=y^0=0$. Using Laplace-method also for the inhomogeneous part, we obtain for the full analytic solution:
\begin{claim}
\begin{align}
G^F(x^0,y^0,\mathbf{P};\mathbf{r},\mathbf{r}^\prime) = G^F_h(x^0,y^0,\mathbf{P};\mathbf{r},\mathbf{r}^\prime) + G^F_m(x^0,y^0,\mathbf{P};\mathbf{r},\mathbf{r}^\prime) \;, \label{eq:fullPNREFT}
\end{align}
where the homogeneous part is given by
\begin{align}
G^F_h(x^0,y^0,\mathbf{P};\mathbf{r},\mathbf{r}^\prime) = \int \text{d}^3 r_1 \text{d}^3 r_2 G^\rho(x^0,\mathbf{P};\mathbf{r},\mathbf{r}_1) G^F_h(0,0,\mathbf{P};\mathbf{r}_1,\mathbf{r}_2) [G^\rho(y^0,\mathbf{P};\mathbf{r}^\prime,\mathbf{r}_2)]^\dagger \;, \label{eq:exacthom}
\end{align}
and $G^F_m$ is independent of the initial condition and referred to as a memory term, given by:
\begin{align}
G^F_m(x^0,y^0,\mathbf{P};\mathbf{r},\mathbf{r}^\prime)= \int_0^{x^0} \text{d}t_1 \int_0^{y^0} \text{d}t_2  G^\rho(x^0-t_1,\mathbf{P};\mathbf{r},0) \Gamma^F(t_1-t_2,\mathbf{P}) G^\rho(t_2-y^0,\mathbf{P};0,\mathbf{r}^\prime) \;.
\end{align}
\end{claim}
One can see that only $s$-wave states contribute to the memory term, while the homogeneous solution also covers the free solution of the non-s-wave states. With this result one can study coherence effects from the initial condition, that will be presented elsewhere. Here, we focus in the following on special cases and limits to recover our earlier gradient-method results.

\subsubsection{Diagonal density matrix and large time limit}
From Eq.~\eqref{eq:fullPNREFT}, one obtains that the initial value of the homogeneous part equals $G^F$ at $x^0=y^0=0$:
\begin{align}
G^{F}(x^0=0,y^0=0,\mathbf{P};\mathbf{r},\mathbf{r}^\prime)=G_h^F(0,0,\mathbf{P},\mathbf{r},\mathbf{r}^\prime).
\end{align}
With this relation, we can obtain the initial condition from the anti-commutator of $G^{+-},G^{-+}$ (definition of $G^{F}$) at initial times. Assuming the initial density matrix is diagonal in momentum space (that neglects coherence), we get, consistent with the differential equations:
\begin{align}
G_h^F(0,0,\mathbf{P},\mathbf{r},\mathbf{r}^\prime) = \sumint\limits_\lambda \psi_\lambda(\mathbf{r}) \psi^\star_\lambda(\mathbf{r}^\prime) \bigg[ \frac{1}{2} + f_\lambda(\mathbf{P}) \bigg] \;. \label{eq:inits}
\end{align}
To compute the spectral function in the homogenous part \eqref{eq:exacthom}, one can realize that the time argument is always positive, implying that the spectral function can be conveniently computed from the retarded correlator:
\begin{align}
G^\rho(x^0,\mathbf{P};\mathbf{r},\mathbf{r}_1)= G^R(x^0,\mathbf{P};\mathbf{r},\mathbf{r}_1) \text{ for } x^0>0.
\end{align}
It is known (see, e.g., \cite{Beneke:2017rdn,PhysRevC.6.114}), that the retarded correlation function for non-hermitian potentials can be computed from a bi-orthogonal set as:
\begin{align}
G^R(\omega,\mathbf{P};\mathbf{r}, \mathbf{r}^\prime) = \sumint\limits_\lambda \frac{i \psi_\lambda(\mathbf{r})\tilde{\psi}^\star_\lambda(\mathbf{r}^\prime)}{\omega -\mathbf{P}^2/(4m)-E_\lambda + i \epsilon} \;. \label{eq:retardbiorth}
\end{align}
The energy $E_\lambda$ is always real for scattering states, while complex for s-wave bound states. For non-hermitian potentials, the usual $\psi, \psi^\star$ set is incomplete, implying also that there is no delta function on the right hand side in \eqref{eq:inits} for the vacuum part.

The wave function solutions can be obtained from the Lippmann-Schwinger method, given by:
\begin{align}
\psi_\mathbf{p}(\mathbf{r}) &= \psi^0_\mathbf{p}(\mathbf{r}) - \Gamma^R(P) \frac{G^R_0(E;\mathbf{r},0) \psi^0_\mathbf{p}(0)}{1+ \Gamma^R(P)  G^R_0(E;0,0)} \;, \\
\tilde{\psi}_\mathbf{p}(\mathbf{r}) &= \psi^0_\mathbf{p}(\mathbf{r}) + [\Gamma^R(P)]^\star \frac{G^R_0(E;\mathbf{r},0) \psi^0_\mathbf{p}(0)}{1- [\Gamma^R(P) ]^\star G^R_0(E;0,0)} \;,\\
\psi_\lambda(\mathbf{r}) &= - \Gamma^R G^R_0(E=E_\lambda;\mathbf{r},0) \psi_\lambda(0) \label{eq:bound1}\;,\\
\tilde{\psi}_\lambda(\mathbf{r}) &= + [\Gamma^R ]^\star G^R_0(E=\tilde{E}_\lambda;\mathbf{r},0) \tilde{\psi}_\lambda(0) \label{eq:bound2} \,.
\end{align}
The complex Eigenvalues of the bound states are formally defined from the solution of~\footnote{
In a concrete example discussed in Section~\ref{sec:example} it turns out that for non-relativistic systems, one can neglect the energy dependence and treat $\Gamma^R$ as a constant. In order for the completeness relation to hold, this has to be consistently done for scattering and bound states. At the leading order  $\Im E_n=- \Re{\Gamma^R} |\psi^0_n(0)|^2 $.}:
\begin{align}
1 + \Gamma^R(\omega_\lambda,\mathbf{P}) G^R_0(\omega_\lambda,\mathbf{P};0,0)=0\;, \quad 1 - [\Gamma^R(\tilde{\omega}_\lambda,\mathbf{P})]^\star G^R_0(\tilde{\omega}_\lambda,\mathbf{P};0,0)=0. \label{eq:complexE}
\end{align}
Notice that $\Im E_\lambda  \leq 0$ must hold due to the $i\epsilon$ boundary condition in \eqref{eq:retardbiorth}. Having the complex Eigenvalues, and inserting them into \eqref{eq:bound1} and \eqref{eq:bound2}, the bound state wavefunctions must be normalized to unity as:
\begin{align}
\int \text{d}^3 r \psi_\lambda(\mathbf{r}) \tilde{\psi}^\star_\lambda(\mathbf{r}) =1.
\end{align}
In this way, the bi-orthogonal set satisfies the completeness relation:
\begin{align}
\sumint\limits_\lambda \psi_\lambda(\mathbf{r}) \tilde{\psi}^\star_\lambda(\mathbf{r}^\prime) = \sumint\limits_\lambda \tilde{\psi}_\lambda(\mathbf{r}) \psi_\lambda^\star(\mathbf{r^\prime})  = \delta(\mathbf{r}-\mathbf{r}^\prime)\;.
\end{align}
With this relation, one can verify that \eqref{eq:retardbiorth} indeed solves the retarded equation. 

Using the bi-orthogonal spectral representation of the retarded Green's function in \eqref{eq:retardbiorth} has the advantage that the Fourier transform is an elementary integral, given by:
\begin{align}
G^{\rho}(t,\mathbf{P};\mathbf{r},\mathbf{r}^\prime)=G^{R}(t,\mathbf{P};\mathbf{r},\mathbf{r}^\prime)=\sumint\limits_\lambda \psi_\lambda(\mathbf{r}) \tilde{\psi}^\star_\lambda(\mathbf{r}^\prime) e^{-i E_{\mathbf{P},\lambda}t} \text{ for } t>0.
\end{align}
With this, the radial integrals in the homogenous part are easy to evaluate, as the bi-orthogonal set also satisfies:
\begin{align}
\int \text{d}^3 r \psi_\lambda(\mathbf{r}) \tilde{\psi}^\star_{\lambda^\prime} (\mathbf{r}) =\delta_{\lambda \lambda^\prime}\;.
\end{align}
Using this and performing the radial integration, we obtain for the homogeneous part:
\begin{align}
G^F_h(x^0,y^0,\mathbf{P};\mathbf{r},\mathbf{r}^\prime) = \sumint\limits_\lambda e^{\Im(E_\lambda) (x^0+y^0)}e^{-i \Re(E_{\mathbf{P},\lambda})(x^0-y^0) }\psi_\lambda(\mathbf{r})\psi^\star_\lambda(\mathbf{r}^\prime) \bigg[ \frac{1}{2} + f_\lambda(\mathbf{P}) \bigg] \;.
\end{align}
Notice that $\tilde \psi$ disappeared. In the large time $T$ limit, we obtain in Fourier space:
\begin{align}
& \int_{-2T}^{2T} \text{d}t e^{i \omega t} G^F_h(T+t/2,T-t/2,\mathbf{P};\mathbf{r},\mathbf{r}^\prime) |_{T(E-\text{Re}E_\lambda) \gg 1} = \\ &\sum\limits_\mathcal{B} e^{- \Gamma_n \delta_{\mathcal{B} ,n} T} (2\pi) \delta(E - \Re E_\mathcal{B} )\psi_\mathcal{B} (\mathbf{r})\psi^\star_\mathcal{B} (\mathbf{r}^\prime) \bigg[ \frac{1}{2} + f_\mathcal{B}(\mathbf{P}) \bigg] \nonumber \\
+ & \int \frac{\text{d}^3q}{(2\pi)^3} (2\pi) \delta(E - \mathbf{q}^2/m )\psi_\mathbf{q} (\mathbf{r})\psi^\star_\mathbf{q} (\mathbf{r}^\prime) \bigg[ \frac{1}{2} + f_\mathbf{q}(\mathbf{P}) \bigg]\;. \nonumber
\end{align}
The s-wave decay width is defined from the imaginary part of the complex Eigenvalue $\Gamma_n \equiv -2\Im E_n $. All bound states are on-shell and the scattering state part is time translational invariant, consistent with our earlier results obtained from the gradient method. Only $s$-wave bound states decay exponentially, but on-shell. For our target component at the origin, this implies in the large $T$ limit:
\begin{claim}
\begin{align}
G^F_h(T,\omega,\mathbf{P};0,0) &=\sum_n e^{-\Gamma_n T} (2\pi) \delta(E-\Re E_n) |\psi_n(0)|^2 \bigg[ \frac{1}{2} + f_n(\mathbf{P}) \bigg] \label{eq:resulthom}\\
&+ \theta(E) \frac{m^2}{2\pi} \sqrt{\frac{E}{m}} \frac{S(v)}{|1+ \Gamma^R(P) G^R_0(E;0,0)|^2} \bigg[ \frac{1}{2} + f_\mathbf{p}(\mathbf{P}) \bigg]\;. \nonumber
\end{align}
\end{claim}
The Sommerfeld-enhanced scattering states respect partial-wave unitarity and are stationary. The unitarization is now temperature dependent due to $\Gamma^R$. This result agrees for the scattering states with the NREFT approach using the gradient method in Section~\ref{sec:NRpaircreation}. However, as we will show in a concrete example, this correction is exponentially suppressed for non-relativistic systems and to leading order $\Gamma^R$ can be approximated to be $P$ independent, which simplifies also the self-consistent computation of the bound states. Notably, the energy gap between bound states remains, even though the spectral function is continuous.

We turn to the memory term and rewrite it as:
\begin{align}
&G^F_m(x^0,y^0,\mathbf{P};\mathbf{r},\mathbf{r}^\prime)= \\
&\int_{-\infty}^\infty \frac{\text{d}\omega}{(2\pi)} \Gamma^F(\omega,\mathbf{P}) e^{-i\omega(x^0-y^0)} \int_0^{x^0} \text{d}y_1 G^\rho(y_1,\mathbf{P},\mathbf{r},0) e^{i\omega y_1}  \int_0^{y^0} \text{d}y_2 G^\rho(-y_2,\mathbf{P},0,\mathbf{r}^\prime) e^{-i\omega y_2} \;. \nonumber
\end{align}
As $x^0,y^0>0$, it follows that $y_1$ and $y_2$ are positive. For these time arguments it follows that the spectral function can be expressed in terms of the retarded correlator as:
\begin{align}
G^\rho(y_1,\mathbf{P},\mathbf{r},0)&=G^R(y_1,\mathbf{P},\mathbf{r},0) \;, \\
G^\rho(-y_2,\mathbf{P},0,\mathbf{r}^\prime)&=-G^A(-y_2,\mathbf{P},0,\mathbf{r}^\prime) = [G^R(y_2,\mathbf{P},\mathbf{r}^\prime,0)]^\star \;.
\end{align}
Taking the large time limit $x^0,y^0 \rightarrow \infty$ in the integrals over the retarded correlator we get for our target component at the origin and in Fourier space of the time difference variable:
\begin{align}
G^F_m(T,\omega ,\mathbf{P};0,0) = \big| G^R(E;0,0) \big|^2 \Gamma^F(\omega,\mathbf{P})\;.
\end{align}
To further simplify this, we consider a concrete example of bosons in the equilibrium environment, discussed in Section~\ref{sec:NRpaircreation} and Appendix \ref{app:example}. For this, the KMS relation $\Gamma^{+-}= e^{-\beta(2m + \omega)} \Gamma^{-+}$ holds, implying the spectral function representation: 
\begin{align}
\Gamma^F(\omega,\mathbf{P})=\frac{1}{2} \coth[\beta(2m + \omega)/2] \Gamma^\rho(\omega,\mathbf{P}) = \bigg[ \frac{1}{2} + f_B^\text{eq}(2m + \omega)\bigg] \Gamma^\rho(\omega,\mathbf{P}) \;.
\end{align}
Noting that the squared retarded correlator can be written as
\begin{align}
| G^R(E;0,0)|^2 &= \frac{|G^R_0(E;0,0)|^2}{|1+\Gamma^R(P) G^R_0(E;0,0)  |^2} \label{eq:rewriting}\\ &= \frac{1}{2\Im i [\Gamma^R(P)]^\star} \bigg[ G^\rho(E;0,0) - \frac{G^\rho_0(E;0,0)}{| 1+\Gamma^R(P) G^R_0(E;0,0)  |^2} \bigg] \;, \nonumber
\end{align}
where we used
\begin{align}
G^\rho(E;0,0) &= 2 \Im i \frac{G^R_0(E;0,0)(1+[\Gamma^R(P)]^\star [G^R_0(E;0,0)]^\star)}{| 1+\Gamma^R(P) G^R_0(E;0,0)  |^2} \\ &=\frac{G^\rho_0(E;0,0)+ 2\Im i[\Gamma^R(P)]^\star|G^R_0(E;0,0)|^2}{| 1+\Gamma^R(P) G^R_0(E;0,0)|^2} \;, \nonumber
\end{align}
the memory term can be written as:
\begin{claim}
\begin{align}
G_m^F(T,\omega,\mathbf{P};0,0) =  \frac{1}{2} \coth[\beta(2m + \omega)/2] \bigg[ G^\rho(E;0,0) - \frac{G_0^\rho(E;0,0)}{| 1+\Gamma^R(P) G^R_0(E;0,0)|^2} \bigg] \text{ , for } T \rightarrow \infty.
\end{align}
\end{claim}
The second term in the parenthesis vanishes for bound states, due to the same reason as discussed in Section~\ref{sec:integral}. Thus, bound states satisfy the KMS relation for large time $T$ as the homogeneous solution exponentially decays, consistent with the dissipation-fluctuation theorem which requires full $G^\rho$ (first term in parenthesis). For scattering states, the second term in the parenthesis does not vanish and violates in general the KMS condition. However,
assuming thermal equilibrium for the scattering states in the homogeneous solution \eqref{eq:resulthom}, in particular a Bose-Einstein distribution $f_B^\text{eq}$ with zero chemical potential, then this term cancels, and the KMS relation is satisfied. Note that the dynamics of the scattering states is governed by the evolution equation for the two-point function, and this stationary solution obtained here is nothing but the unitarization of the Sommerfeld effect including pair production processes. The latter are responsible that we see steady state solutions for the bound states, which correspond to the thermal attractor solution.

\section{Including pair creation in the number density equation}
\label{sec:NRpaircreation}

We turn back to the NREFT approach to include also the reverse process of pair creation from the thermal environment, relevant to achieve chemical equilibration in the number density equation. Our starting point is the NREFT action in Eq.~\eqref{eq:NRaction}, but now with the general form of the ``matching coefficient" on the CTP contour in Eq.~\eqref{eq:generalmatching}. From this, we derive the particle number density evolution equation:
\begin{equation}
    \dot{n}_\eta = - \sum_\sigma \sigma \int d^4 z \; \Gamma^{+ \sigma} (x,z) G_{\eta \xi}^{\sigma \sigma ++} (z,z,x,x) - \Gamma^{\sigma +} (z, x) G_{\eta \xi}^{++ \sigma \sigma} (x,x,z,z)\;,
\end{equation}
where $\sigma \in \{+,-\}$. To leading order in the gradient expansion, we can express this equation in Wigner space as
\begin{equation}
    \dot{n}_\eta = \int \frac{d^4 P}{(2 \pi)^4} \Gamma^{+-} (P) G_{\eta \xi}^{--++} (T,P; 0, 0) - \Gamma^{-+} (P) G_{\eta \xi}^{++--} (T,P; 0, 0).
    \label{eq:numberdensity}
\end{equation}
In thermal equilibrium, we can see that, due to the KMS relation, both terms cancel and we get a vanishing rate, as expected. In the following, we shall first compute the four-point correlators in the NREFT approach based on the gradient method, compare them to the analytic solutions in the pNREFT approach, and give a concrete example of how to match a specific case. 

\subsection{Self-consistent four-point functions from NREFT}
\label{sec:NRgradientall}

 To self-consistently compute $G_{\eta \xi} (P; 0, 0)$, entering the number density equation, we derive the closed NREFT integral equation of motion for the four-point correlation functions in Appendix~\ref{app:NREFT}. Assuming that the solution $G_{\eta \xi,0} (P; 0, 0)$ with the static potential interactions is known, then to leading order in the gradient expansion we obtain a Bethe-Salpeter type of equation for the four-point correlation functions including the effect of $\Gamma$ (see Eq.~\eqref{eq:BS_static} for a derivation):
\begin{equation}
    G_{\eta \xi}^{\sigma_1 \sigma_1 \sigma_2\sigma_2} (P; 0,0) = G_{\eta \xi,0}^{\sigma_1 \sigma_1\sigma_2 \sigma_2} (P; 0,0) -  \sum_{\sigma_3, \sigma_4} \sigma_3 \sigma_4 G_{\eta \xi,0}^{\sigma_1 \sigma_1  \sigma_3  \sigma_3} (P; 0, 0) \Gamma^{\sigma_3 \sigma_4} (P) G_{\eta \xi}^{\sigma_4 \sigma_4 \sigma_2\sigma_2} (P; 0,0).
\end{equation}
This is a closed matrix equation in the contour indices. We find for the self-consistent solution:
\begin{align}
    G_{\eta \xi}^{\sigma_1 \sigma_1\sigma_2 \sigma_2}(P;0,0) =& \big[ 1 + (\Gamma^{++} G_{\eta \xi,0}^{++++} - \Gamma^{--} G_{\eta \xi,0}^{----} - \Gamma^{+-} G_{\eta \xi,0}^{--++} - \Gamma^{-+} G_{\eta \xi,0}^{++--}) \nonumber \\ + & (\Gamma^{++} \Gamma^{--} - \Gamma^{+-} \Gamma^{-+})(G_{\eta \xi,0}^{++++} G_{\eta \xi,0}^{----} - G_{\eta \xi,0}^{++--} G_{\eta \xi,0}^{--++}) \big]^{-1} \nonumber \\
     \times & \left[\begin{pmatrix}
        G_{\eta \xi,0}^{++++} & G_{\eta \xi,0}^{++--} \\
        G_{\eta \xi,0}^{--++} & G_{\eta \xi,0}^{----}
    \end{pmatrix} + (G_{\eta \xi,0}^{++++} G_{\eta \xi,0}^{----} - G_{\eta \xi,0}^{++--} G_{\eta \xi,0}^{--++}) \begin{pmatrix}
        \Gamma^{--} & \Gamma^{+-} \\
        \Gamma^{-+} & \Gamma^{++}
    \end{pmatrix} \right] \nonumber \\
    =& \frac{1}{\big|1 + \Gamma^R(P) G_{\eta \xi,0}^R(P;0,0)\big|^2} \nonumber \\ \times & \left[\begin{pmatrix}
        G_{\eta \xi,0}^{++++}(P;0,0) & G_{\eta \xi,0}^{++--}(P;0,0) \\
        G_{\eta \xi,0}^{--++}(P;0,0) & G_{\eta \xi,0}^{----}(P;0,0)
    \end{pmatrix} + \big|G_{\eta \xi,0}^R(P;0,0)\big|^2 \begin{pmatrix}
        \Gamma^{--}(P) & \Gamma^{+-} (P)\\
        \Gamma^{-+}(P) & \Gamma^{++}(P) 
    \end{pmatrix} \right]. \label{eq:fullsolutionNREFT}
\end{align}
The relevant components of the second matrix cancel exactly when inserting them into the number density evolution equation, while the first matrix can in general be out of equilibrium. Inserting \eqref{eq:fullsolutionNREFT} into the number density evolution equation \eqref{eq:numberdensity}, we obtain our final result:
\begin{claim}
\begin{equation}
    \dot{n}_{\eta} = \int \frac{d^4 P}{(2 \pi)^4} \frac{1}{\big|1 + \Gamma^R(P) G_{\eta \xi,0}^R (P; 0, 0)\big|^2} \bigg[\Gamma^{+-} (P) G_{\eta \xi,0}^{--++} (P; 0, 0) - \Gamma^{-+} (P) G_{\eta \xi,0}^{++--} (P; 0, 0) \bigg]. \label{eq:numberfull}
\end{equation}
\end{claim}
The common denominator unitarizes Sommerfeld-enhanced annihilation and pair creation for scattering states, which is now in general temperature dependent due to the retarded $\Gamma^R$.

Let us assume that kinetic equilibrium is maintained and write down a Kadanoff-Baym Ansatz for the statistical correlators in the dilute limit:
\begin{align}
G_{\eta \xi,0}^{++--} (P; 0, 0) &= \left(\frac{n_s}{n_s^\text{eq}}\right)^2 e^{-\beta(2m + E+\mathbf{P}^2/(4m))} \theta(E) \frac{m^2}{2\pi} \sqrt{\frac{E}{m}} S(v) \nonumber \\ &+ \sum_n  \frac{n_n}{n_n^\text{eq}} e^{-\beta(2m+E_n+\mathbf{P}^2/(4m))} (2\pi) \delta(E-E_n) |\psi_n(0)|^2 \;, \\
G_{\eta \xi,0}^{--++} (P; 0, 0) &= \theta(E) \frac{m^2}{2\pi} \sqrt{\frac{E}{m}} S(v) \nonumber + \sum_n (2\pi) \delta(E-E_n) |\psi_n(0)|^2 \;.
\end{align}
As $G^R_{\eta \xi,0}(E\sim E_n;0,0)\propto (E-E_n+ i \epsilon)^{-1} + \text{regular}$, the bound states entirely drop out of the collision term in Eq.~\eqref{eq:numberfull} and one obtains the standard Boltzmann equation for scattering states:
\begin{align}
\dot{n}_s = - \langle \sigma v \rangle \left( n_s^2 -  n_{s,\text{eq}}^2 \right)
\end{align}
with standard thermal average of the unitarized Sommerfeld-enhanced annihilation cross section in the dilute limit consistent with~\cite{Blum:2016nrz}.

As we have already discussed, the leading order gradient result in \eqref{eq:fullsolutionNREFT}, gives the stationary solution. For bound states in $G_{\eta \xi}^{++--}$ and $G_{\eta \xi}^{--++}$ in Eq.~\eqref{eq:fullsolutionNREFT} only the Breit-Wigner shapes multiplying $\Gamma^{+-}$ and $\Gamma^{-+}$, respectively, are non-zero, however, as these $\Gamma$ terms cancel in the collision term there is no contribution of stationary bound state solutions. To get the correct equation, one has to solve the time dependent problem, which is the only solution for bound states, in contrast to scattering states, where the only solution is the stationary one (when assuming no coherence effects).

We show in the following that the NREFT and pNREFT results agree for the scattering states. Based on this, one can expect that the time evolution of the bound states matches also in both approaches. In this case, the time evolution of the bound states is included in the analytic solution in Eq.~\eqref{eq:fullPNREFT}.

\subsection{Comparison with pNREFT solution}
We equate the two-time NREFT four-point correlators with the pNREFT four-point correlators as $G_{\eta\xi}(t,\mathbf{x}_1,\mathbf{x}_2;t^\prime,\mathbf{x
}_3,\mathbf{x}_4)=G(x,y;\mathbf{r},\mathbf{r}^\prime) $  (cf., similarly in Ref.~\cite{Brambilla:2017zei}). At LO in the gradient expansion of the number density equation, only $G^{+-}$ and $G^{-+}$ are required. Taking the analytic solution of these from Section \ref{sec:exact} in the large time $T$ limit, where scattering states are stationary, we get
\begin{align}
G^{+-}(T,P;0,0) &= G^F(T,P;0,0) - \frac{1}{2} G^\rho(P;0,0) \\ &= \frac{G^\rho_0(E;0,0)}{|1+ \Gamma^R(P) G_0^R(E;0,0) |^2} \big[ f_\mathbf{p}(\mathbf{P}) - f_B^\text{eq}(2m+\omega)] + f_B^\text{eq}(2m+\omega) G^\rho(P;0,0)\;, \nonumber\\ 
G^{-+}(T,P;0,0) &= G^F(T,P;0,0) + \frac{1}{2} G^\rho(P;0,0) \\ &= \frac{G^\rho_0(E;0,0)}{|1+ \Gamma^R(P) G_0^R(E;0,0) |^2} \big[ (1 + f_\mathbf{p}(\mathbf{P})) - (1+f_B^\text{eq}(2m+\omega))] + (1 + f_B^\text{eq}(2m+\omega)) G^\rho(P;0,0) \;, \nonumber
\end{align}
 where the out-of-equilibrium pair distribution function is on-shell, i.e. $E=\mathbf{p}^2/m$. This is indeed consistent with the NREFT solution in Eq.~\eqref{eq:fullsolutionNREFT}, which are based on the gradient method, when applying the KMS relation $\Gamma^{+-}= e^{-\beta(2m + \omega)} \Gamma^{-+}$ and using the rewriting in Eq.~\eqref{eq:rewriting}. Inserting into the number density equation we obtain:
\begin{align}
    \dot{n}_\eta &= \int \frac{d^4 P}{(2 \pi)^4} \Gamma^{+-} (P) G^{-+}(T,P; 0, 0) - \Gamma^{-+} (P) G^{+-}(T,P; 0, 0) \label{eq:numberpnrfull} \\
    &= \int \frac{d^4 P}{(2 \pi)^4} \frac{G^\rho_0(E;0,0)}{|1+ \Gamma^R(P) G_0^R(E;0,0) |^2} \bigg[ \Gamma^{+-} (P) (1 + f_\mathbf{p}(\mathbf{P})) - \Gamma^{-+} (P) f_\mathbf{p}(\mathbf{P})   \bigg]\;. \nonumber
\end{align}
All equilibrium parts drop out of the collision term under the KMS relation $\Gamma^{+-}= e^{-\beta(2m + \omega)} \Gamma^{-+}$. As in the NREFT approach, the unitarization of the Sommerfeld effect is now temperature dependent in general, and we consider a concrete example next. Note that in kinetic equilibrium, the dilute limit and the molecular chaos assumption, the two-particle phase space distribution is normalized as:
\begin{align}
f_\mathbf{p}(\mathbf{P}) = \frac{n_\eta n_\xi}{n_\eta^\text{eq} n_\xi^{\text{eq}}} e^{-\beta (2m + P^0)}\;.
\end{align}
By change of variables $E=P^0 - \mathbf{P}^2/(4m)$ one obtains the expected form of a Boltzmann equation with standard thermal average in the dilute limit. Thus, the NREFT and pNREFT approach deliver identical results for the scattering states.

\subsection{Example}
\label{sec:example}
 Consider a heavy complex scalar $\Phi$ with mass $m$ that can annihilate into light complex scalar $\phi$ via contact interaction $\mathcal{L}\supset - \lambda \Phi^\dagger \Phi \phi^\dagger \phi$. We integrate out the light degrees of freedom $\phi$ in Appendix~\ref{app:example} and match it to the NR effective action in \eqref{eq:NRaction}. We define $\Gamma$ from the one-loop self-energy $\Im i \Pi$ in Eq.~\eqref{eq:gammaexample} in the non-relativistic limit of $\Phi$, giving:
\begin{align}
&\Gamma^{+-}(P) = (\sigma v)_0 f_B^\text{eq}(m+P^0/2)f_B^\text{eq}(m+P^0/2)\;, \quad 
\Gamma^{-+}(P) = (\sigma v)_0 \left(1+f_B^\text{eq}(m+P^0/2)\right)^2 \label{eq:pmmp} \;,\\
&\Gamma^R(P)= \frac{1}{2} (\sigma v)_0 \left(1+ 2 f_B^\text{eq}(m+P^0/2) \right) \label{eq:gamR} \;.
\end{align}
Inserting \eqref{eq:pmmp} into the number density equation \eqref{eq:numberfull} with NREFT correlators or \eqref{eq:numberpnrfull} with pNREFT correlators, one observes the expected Bose-enhancement factors for annihilation, and the quadratic $f_B^\text{eq}$ dependence for pair creation. Eq.~\eqref{eq:gamR} contains a temperature dependent correction to the unitarization of the Sommerfeld effect\footnote{The structure $1+2f_B^\text{eq}$ can be explained from $\Gamma^{R}=\Gamma^{++} - \Gamma^{+-}$ which is proportional to $(1+f_B^\text{eq})^2 - (f_B^\text{eq})^2=1+2f_B^\text{eq}$, showing that it is the net Bose-enhancement from annihilation and pair production.}:
\begin{align}
\Gamma^R(P) G_{\eta \xi,0}^R (P; 0, 0)=\Gamma^R(P) G^R_0 (P; 0, 0) = \frac{(\sigma v)_0 S(v)\left(1+ 2 f_B^\text{eq}(m+P^0/2) \right) }{4(\sigma v)_\text{uni}} + \text{finite imaginary} \;.
\end{align}
As $f_B^\text{eq}>0$, the resonant Sommerfeld effect will saturate for larger $v$ than in vacuum, consistent with partial-wave unitarity. Notice however, that $f_B^\text{eq} \ll 1$ in the non-relativistic regime of $\Phi$, and therefore it is only a very small temperature dependent correction to the already known result~\cite{Blum:2016nrz}.

$\Gamma^R(P)$ also enters in the complex Eigenvalue equation \eqref{eq:complexE} for the bound states. With energy dependence this is a complicated task. To good approximation, one can take it to be a constant, $\Gamma^R(P) \approx  (\sigma v)_0 \left(1+ 2 e^{-\beta m} \right)/2$, or even neglecting the exponentially suppressed temperature dependent piece. This must be consistently done for scattering and bound states to ensure the completeness relation. Note, however, that the $P^0$ dependence in $\Gamma^{+-/-+}$ must be kept, otherwise the collision term in \eqref{eq:numberdensity} would diverge (in particular the pair production term).

\section{Summary and conclusion}
\label{sec:conclusion}

Cross sections computed from potentials that parametrically admit near-threshold bound states require particular care in order to remain consistent with partial-wave unitarity. This issue is especially relevant in parameter scans of dark matter models, where resonance phenomena can arise in the solution of multi-dimensional Schrödinger equations. In the dark matter literature, the unitarization of the Sommerfeld-enhanced annihilation cross section for scattering states in vacuum was achieved by incorporating the annihilation potential self-consistently in the Schrödinger equation~\cite{Blum:2016nrz}. This framework was later extended to multiple channels and arbitrary partial waves~\cite{Parikh:2024mwa}; see also~\cite{Braaten:2017dwq, Flores:2024sfy, Watanabe:2025kgw, Flores:2025uoh}.

In this work, we revisited the simplest case consisting of a single channel and the $\ell=0$ partial-wave contribution to particle number changing reaction. Our primary objective was to determine how bound-state properties are modified when pair annihilation and creation are treated self-consistently, and in particular whether bound states contribute as on-shell degrees of freedom in chemical equilibration or instead acquire an intrinsically off-shell character due to the finite decay width induced by annihilation.  
One of the main results of this work is to clarify this aspect.

To this end, we employed NREFT and pNREFT methods and adapted those to the Keldysh-Schwinger formalism. We restricted the analysis to the minimal setup that already captures the essential physics, namely a static potential together with the pair annihilation/creation operator. In this framework, we showed that a self-consistent treatment of pair annihilation and creation requires the computation of four-point correlation functions on the closed-time-path contour.

Within pNREFT, we derived the solutions of the  four-point correlation functions in two complementary ways: the first is based on the gradient method in Section~\ref{sec:integral}, and the second through an analytic solution for an initial value problem in Section~\ref{sec:exact}. The two approaches deliver consistent results in expected limits. The resulting solutions exhibit a clear qualitative difference between scattering and bound states. Scattering states admit stationary solutions, but their chemical evolution is not determined by the four-point function alone and requires, in addition, the two-point equation for the particle number density. Bound states behave differently: their Boltzmann equations emerge dynamically from the four-point correlator itself under semi-classical assumptions and chemical equilibration occurs in the large time limit.

For scattering states, the self-consistent stationary solution provides the unitarization of the Sommerfeld enhancement, which was obtained in both NREFT and pNREFT approaches. Once pair creation is included, this unitarization becomes temperature dependent, although the corresponding thermal corrections remain small in the non-relativistic regime. It can be concluded that earlier vacuum results for the scattering states are reliable for dark matter freeze-out studies.

For bound states, our main result is that the self-consistent treatment leads to an on-shell contribution to the out-of-equilibrium decay dynamics. This conclusion is non-trivial, since the corresponding spectral functions are continuous and exhibit a Breit-Wigner profile as a consequence of the finite width. The on-shell chemical equilibration is in agreement with analogous results for unstable elementary particles~\cite{Anisimov:2008dz}. Here, the self-consistent treatment of bound states does, however, give a correction to the wave function at the origin and, accordingly, the bound-state decay rates.

Our main analytic result is the pNREFT solution in Eq.~\eqref{eq:fullPNREFT}, which also incorporates coherence effects. A numerical analysis of this solution is left for future work.

\acknowledgments

We would like to thank Kallia Petraki, Marco Drewes, Mathias Garny, Shuo Fang, Stefan Lederer, Tom Margosch, Tracy Slatyer, Xiaojun Yao, and Yukinao Akamatsu for discussion.

\appendix

\section{Matching the example}
\label{app:example}

We give an example of deriving the real part of the matching coefficient $\Gamma$ for the simplest case of contact annihilation of charged scalars, given by the model
\begin{align}
\mathcal{L} \supset - \lambda \Phi^\dagger \Phi \phi^\dagger \phi \;,\label{eq:toyann}
\end{align}
where $\Phi$ is the heavy field of mass $m$ and $\phi$ is massless and constitutes (part of) the environment.

The derivation is relatively straightforward and follows closely the familiar zero-temperature analysis. More generally, we work in a weakly coupled theory and assume a factorized initial density matrix for the system and environment, so that the environmental degrees of freedom can be integrated out using standard methods. We therefore keep the discussion brief up to the stage where pair creation becomes relevant. At that point, however, the conventional vacuum derivation requires a suitable extension, since a description of chemical equilibration necessitates retaining exponentially suppressed terms in the non-relativistic effective action that are usually neglected.

In deriving the non-relativistic effective action, we perform a field split of the heavy field, where the non-relativistic component reads:
\begin{align}
\Phi_\text{NR}(x) = \frac{1}{\sqrt{2m}} \eta(x) e^{-imx^0} + \frac{1}{\sqrt{2m}} \xi(x) e^{i mx^0} \;,
\end{align}
where $\eta$ annihilates a particle while $\xi$ creates an anti-particle. We integrate out the light degrees of freedom ($\phi$) at the one-loop level with self-energy given by:
\begin{align}
\Pi(x-y)=D(x-y)D(y-x)\;, \quad D(x,y)=\langle T_{\mathcal{C}} \phi(x) \phi^\dagger(y) \rangle \;. \label{eq:defmodelselfenergy}
\end{align}
 In terms of the above quantities, the effective action is proportional to contour integrals over
\begin{align}
&\Phi_\text{NR}^\dagger(x) \Phi_\text{NR}(x) \Pi(x-y) \Phi_\text{NR}^\dagger(y)\Phi_\text{NR}(y) = \label{eq:fourscalar}\\ 
& \frac{1}{4m^2} \bigg[ \left(\eta^\dagger(x) \eta(x) +\xi^\dagger(x) \xi(x) \right) \Pi(x-y) \left(\eta^\dagger(y) \eta(y) +\xi^\dagger(y) \xi(y) \right) \bigg] + \nonumber \\
&\frac{1}{4m^2} \bigg[ \eta^\dagger(x) \xi(x) \Pi(x-y) e^{i2m(x^0-y^0)} \xi^\dagger(y) \eta (y) + \xi^\dagger(x) \eta(x) \Pi(x-y) e^{-i2m(x^0-y^0)}  \eta^\dagger(y)  \xi(y)  \bigg] \nonumber \\
&+ ... \nonumber
\end{align}
where $...$ contains a product of four annihilation and a product of four creation operators, which can be omitted as they don't have contractions. The first line after the equality contains elastic reactions as it commutes with particle and antiparticle number density operators. It can be included in the static potential part. The second line contains pair annihilation and creation and is of our interest. In Fourier space, the first oscillatory term is:
\begin{align}
\int\limits_{x^0,y^0 \in \mathcal{C}} \text{d}^4x \text{d}^4y \; \mathcal{O}^\dagger(x) \mathcal{O}(y) \Pi (x-y) e^{i2m(x^0-y^0)} &= \sum_{\sigma_1 \sigma_2} \sigma_1 \sigma_2 \int \frac{\text{d}^4p}{(2 \pi)^4} \; \mathcal{O}^\dagger_{\sigma_1}(p) \mathcal{O}_{\sigma_2}(p) \Pi^{\sigma_1 \sigma_2} (2m + P^0,\mathbf{p})
\end{align}
where $\mathcal{O}=\xi^\dagger \eta $. The second oscillatory term in Eq.~\eqref{eq:fourscalar}, after relabeling $x \leftrightarrow y$ and using Eq.~\eqref{eq:defmodelselfenergy}, gives the same contribution as the first. To get the annihilation cross section in the usual vacuum computation from the self-energy, one makes use of the non-relativistic hierarchy  $m \gg |\mathbf{p}| \gg P^0 \sim p^2/m$ and expands $\Im i \Pi (2m + P^0,\mathbf{p})|_{T=0} \simeq \Im i \Pi (2m, 0) $. However, here on the CTP contour, including pair production and Bose-enhancement effects, requires particular care. In order to derive consistent non-relativistic Boltzmann equations from the self-energy that incorporate both pair annihilation and creation, the standard zero-temperature NREFT counting rules must be appropriately modified to include exponentially small pair production terms as described below.

For $\Im i \Pi^{\sigma_1 \sigma_2}$ with a general energy argument $\omega$ we find (see Eq.~\eqref{eq:resultselfenergy} for derivation):
\begin{align}
& \Im i \Pi^{\sigma_1 \sigma_2}(\omega,\mathbf{p}=0) = \frac{1}{16 \pi} \times \label{eq:selfmainresult} \\ &\bigg[
\begin{pmatrix}
1 & 2 \theta(-\omega) \\
2 \theta(\omega) & 1
\end{pmatrix}
+2 f_B^\text{eq}(|\omega|/2)
\begin{pmatrix}
1 & 2 \theta(-\omega) \\
2 \theta(\omega) & 1
\end{pmatrix}
+2 f_B^\text{eq}(|\omega|/2) f_B^\text{eq}(|\omega|/2)
\begin{pmatrix}
1 & 1 \\
1  & 1
\end{pmatrix}
\bigg]\;. \nonumber
\end{align}
Now, the rule for non-relativistic $\omega = 2m + P^0$ is to neglect $P^0$ everywhere, except in the Bose-Einstein distribution function $f_B^\text{eq}(|\omega|/2) =(e^{\beta |\omega|/2}-1)^{-1}$. As $|P^0|\ll m$ by non-relativistic assumption, one can also neglect the absolute value symbol in the Bose-distribution. We match $\Gamma(x,y)$ in the NR action \eqref{eq:NRaction} in momentum space to this model using these rules:
\begin{align}
&\Gamma^{\sigma_1 \sigma_2}(P^0,\mathbf{p})\equiv \frac{\lambda^2}{4 m^2} \Im i \Pi^{\sigma_1 \sigma_2}(2m + P^0,\mathbf{p}=0) = \frac{1}{2} (\sigma v)_0 \times \label{eq:gammaexample} \\
&\bigg[
\begin{pmatrix}
1 & 0 \\
2 & 1
\end{pmatrix}
+2 f_B^\text{eq}[(2m+P^0)/2]
\begin{pmatrix}
1 & 0 \\
2 & 1
\end{pmatrix}
+2 f_B^\text{eq}[(2m+P^0)/2] f_B^\text{eq}[(2m+P^0)/2]
\begin{pmatrix}
1 & 1 \\
1  & 1
\end{pmatrix}
\bigg] \nonumber
\end{align}
where the tree-level annihilation cross section for $\Phi \Phi \rightarrow \phi \phi$ is:
\begin{align}
(\sigma v)_0 = \frac{\lambda^2}{32 \pi m^2} \;.
\end{align}
Notice that $\Gamma^{++} + \Gamma^{--}= \Gamma^{+-} + \Gamma^{-+}$, and it obeys the KMS relation: $\Gamma^{+-}= e^{-\beta(2m+P^0)}\Gamma^{-+}$. The zero temperature part in momentum and coordinate space is:
\begin{align}
\Gamma^{\sigma_1 \sigma_2}(P^0,\mathbf{p})|_{T=0} = \frac{1}{2} (\sigma v)_0 
\begin{pmatrix}
1 & 0 \\
2 & 1
\end{pmatrix}
\;, \quad \Gamma(x,y)|_{T=0} = \frac{1}{2} (\sigma v)_0 \delta^4(x-y)
\begin{pmatrix}
1 & 0 \\
2 & 1
\end{pmatrix}
\;,
\end{align}
which is a local interaction in time and space. It belongs to pair annihilation only in the dilute regime. The finite temperature pieces account for Bose-enhancement correction for annihilation as well as include pair creation. They lead to a non-local form in time due to the $P^0$ dependence, which is needed to get a consistent non-relativistic Boltzmann equation. In particular, without the $P^0$ dependence in Eq.~\eqref{eq:gammaexample}, which is usually regarded as a sub-dominant contribution, the pair creation part of the collision term in \eqref{eq:numberfull} would diverge.

\paragraph*{Derivation of Eq.~\eqref{eq:selfmainresult}}
For the light degrees of freedom in thermal equilibrium, we use the Feynman rules (see, e.g. Eq. IV.20 in \cite{Lundberg:2020mwu}):
\begin{align}
    D^{+-} (K) =&  \slashed{\delta}(K^2-m^2_\phi) [\theta(-K^0) + f_B^\text{eq}(|K^0|)]\;, \\
    D^{-+} (K) =&  \slashed{\delta}(K^2-m^2_\phi) [\theta(K^0) + f_B^\text{eq}(|K^0|)]\;,\\
    D^{++} (K) =& \frac{i}{K^2 -m^2_\phi + i \epsilon} + \slashed{\delta}(K^2-m^2_\phi)  f_B^\text{eq} (|K^0|) \;, \\
   D^{--} (K) =& \frac{-i}{K^2 -m^2_\phi - i \epsilon} + \slashed{\delta}(K^2-m^2_\phi)  f_B^\text{eq} (|K^0|)\;.
\end{align}
The small mass $m_\phi$ we take to zero in final results. We start with $\Pi^{++}$, $\Pi^{--}$ and introduce superscripts to label vacuum by $v$ and finite temperature contributions by $T$. Starting from the zero temperature contribution: 
\begin{align}
i \Pi^{++/--}_{vv} (P) & = i \tilde{\mu} \int \frac{\text{d}^d k}{(2\pi)^d} \frac{\pm i}{(k+P)^2-m^2_\phi \pm i \epsilon} \frac{\pm i}{k^2-m^2_\phi \pm i \epsilon} \nonumber\\
&= \frac{\pm 1}{(4 \pi)^2} \left( \frac{1}{\epsilon} - \int_0^1 \text{d}x \ln\left( \frac{m^2_\phi-x \bar{x} P^2 \mp i \epsilon}{\mu^2} \right) + \mathcal{O}( \epsilon) \right) \;. \label{eq:Iself}
\end{align}
The overall different sign in the anti-time ordered case originates from the Wick rotation in the opposite direction.
The imaginary part for the massless case is:
\begin{align}
\Im i \Pi^{++/--}_{vv} (P) &= \frac{1}{16\pi^2}\Im \bigg[ \mp \int_0^1 \text{d}x \ln\left( \frac{-x \bar{x} P^2 \mp i \epsilon}{\mu^2} \right) \bigg] = \frac{1}{16\pi} \theta(P^2) \;, \\
\Im i \Pi^{++/--}_{vv}(P^0,\mathbf{p}=0) &= \frac{1}{16 \pi} \;.
\end{align}
We turn to the mixed vacuum and finite temperature contribution:
\begin{align}
\Im i \Pi^{++/--}_{vT}(P) &= \Im i \int \slashed{d}^4k \frac{\pm i}{(P-K)^2-m^2_\phi \pm i \epsilon} \slashed{\delta}(K^2-m^2_\phi) f(|K^0|) \nonumber \\
&=\frac{1}{2} \int \slashed{d}^4k \slashed{\delta}((K-P)^2-m^2_\phi)\slashed{\delta}(K^2-m^2_\phi) f(|K^0|) \nonumber \\
&=\frac{1}{2} \int \slashed{d}^4k \slashed{\delta}(P^2-2KP)\slashed{\delta}(K^2-m^2_\phi) f(|K^0|)\;,\\
\Im i \Pi^{++/--}_{vT}(P^0,0) &= \frac{1}{2} \int \slashed{d}^4k \slashed{\delta}(P_0^2-2K^0P^0)\slashed{\delta}(K^2-m^2_\phi) f(|K^0|) \nonumber \\
&= \frac{1}{2} \int \frac{\slashed{d}^3k}{2 |k|} f(|k|) [ \slashed{\delta}(P_0^2-2|k| P^0) + \slashed{\delta}(P_0^2+2|k| P^0) ] \nonumber  \\
&= \frac{1}{2} \int \frac{\slashed{d}^3k}{2 |k| 2 |P^0|} f(|k|) \slashed{\delta}(|k|-|P_0|/2) \nonumber \\
&= \frac{1}{16 \pi} \frac{1}{e^{\beta|P^0|/2}-1} \;.
\end{align}
$\Im i \Pi^{++/--}_{Tv}(P^0,0) $ gives the same. The purely temperature dependent contribution can conveniently be evaluated from
\begin{align}
\Im i \Pi^{++/--}_{TT}(P) &= \int \slashed{d}^4K_1  \slashed{d}^4K_2 \slashed{\delta}^4(P-K_1-K_2) \slashed{\delta}(K_1^2) \slashed{\delta}(K_2^2) f(|K_1^0|)f(|K_2^0|) \;.
\end{align}
Setting $\mathbf{p}=0$, the momentum delta function enforces $\mathbf{k}_1=-\mathbf{k}_2$ (back-to-back). Now assuming $P^0>0$, then the first delta function requires $K_1^0>0$ and $K_2^0>0$ for a non-vanishing result. Therefore:
\begin{align}
\Im i \Pi^{++/--}_{TT}(P^0>0,0) &= \int \frac{\slashed{d}^3k}{4 |k|^2} \slashed{\delta}(P^0-2|k|) f(|k|) f(|k|) \nonumber \\
&= \frac{1}{8 \pi} \frac{1}{e^{\beta P^0/2}-1} \frac{1}{e^{\beta P^0/2}-1} \;.
\end{align}
Repeating for $P^0<0$ one finds
\begin{align}
\Im i \Pi^{++/--}_{TT}(P^0<0,0) &= \frac{1}{8 \pi} \frac{1}{e^{\beta |P^0|/2}-1} \frac{1}{e^{\beta |P^0|/2}-1} \;.
\end{align}
Adding all contributions, $\Pi = \Pi_{vv} + 2\Pi_{vT} + \Pi_{TT}$, one obtains for general $P^0$:
\begin{align}
\Im i \Pi^{++/--}(P^0,0) = \frac{1}{16 \pi} \left(1 + 2 \frac{1}{e^{\beta|P^0|/2}-1} + 2 \frac{1}{e^{\beta|P^0|/2}-1} \frac{1}{e^{\beta|P^0|/2}-1} \right).
\end{align}
Last, we compute $\Pi^{+-/-+}$, given for general $P^0$ by
\begin{align}
\Im i \Pi^{+-/-+}(P^0,0) = \frac{1}{8 \pi} \bigg[ \theta( \mp P^0) \left(1+2\frac{1}{e^{\beta|P^0|/2}-1} \right) +\frac{1}{e^{\beta|P^0|/2}-1}\frac{1}{e^{\beta|P^0|/2}-1} \bigg] \;.
\end{align}
Collecting all pieces, the final result for the one-loop self-energy for general $P^0$ then reads:
\begin{align}
& \Im i \Pi(P^0,\mathbf{p}=0) = \frac{1}{16 \pi} \times \label{eq:resultselfenergy} \\ &\bigg[
\begin{pmatrix}
1 & 2 \theta(-P^0) \\
2 \theta(P^0) & 1
\end{pmatrix}
+2 \frac{1}{e^{\beta|P^0|/2}-1}
\begin{pmatrix}
1 & 2 \theta(-P^0) \\
2 \theta(P^0) & 1
\end{pmatrix}
+2 \frac{1}{e^{\beta|P^0|/2}-1} \frac{1}{e^{\beta|P^0|/2}-1}
\begin{pmatrix}
1 & 1 \\
1  & 1
\end{pmatrix}
\bigg] \;. \nonumber
\end{align}

\section{Technical details of the pNREFT formalism}
\label{app:freepNR}

\subsection*{Free correlators}
The equations of motion for the CTP correlators without annihilation read:
\begin{align}
(i \partial_{x^0} - h_x)G_0(x,y;\mathbf{r},\mathbf{r}^\prime) &= i \delta^{4}_\mathcal{C}(x-y) \delta(\mathbf{r} -\mathbf{r}^\prime) \label{eq:firsteom1}\\
(-i \partial_{y^0} - h_y^\prime)G_0(x,y;\mathbf{r},\mathbf{r}^\prime) &= i \delta^{4}_\mathcal{C}(x-y) \delta(\mathbf{r} -\mathbf{r}^\prime) \label{eq:secondeom2}
\end{align}
We solve for the statistical $G^{+-}_0$ and retarded correlator $G^{R}_0= G^{++}_0 -G^{+-}_0$. Having the solution of both, all other correlators $-+,++,--$ follow from basic definitions as:
\begin{align}
G^{++}_0 &= G^{R}_0+G^{+-}_0\;, \quad G^A_0 =-(G^{R}_0)^\star \;, \quad  G^{-+}_0 = G^{++}_0 - G^A_0 \;, \quad G^{--}_0 =G^{A}_0 + G_0^{+-} \;.
\end{align}
Let us start with the EoM of the statistical correlator, which are homogeneous:
\begin{align}
(i \partial_{x^0} - h_x)G_0^{+-}(x,y;\mathbf{r},\mathbf{r}^\prime) &= 0 \;, \\
(-i \partial_{y^0} - h_y^\prime)G_0^{+-}(x,y;\mathbf{r},\mathbf{r}^\prime) &= 0 \;.
\end{align}
The statistical correlator is defined in terms of the two-body field and the initial density matrix as:
\begin{align}
G^{+-}(x,y;\mathbf{r},\mathbf{r}^\prime) = \text{Tr}[ \rho(t_0) S^\dagger_H(y^0,\mathbf{y},\mathbf{r}^\prime) S_H(x^0,\mathbf{x},\mathbf{r}) ] \;.
\end{align}
The two-body field in the Heisenberg picture can be expressed in the energy basis as:
\begin{align}
S_H(x^0,\mathbf{x},\mathbf{r})= \int\limits_\mathbf{P} \sumint\limits_{\lambda} e^{i \mathbf{P}\mathbf{x}} \psi_\lambda(\mathbf{r}) e^{-i E_{\mathbf{P},\lambda}(x^0 - t_0)} b_S(\mathbf{P}, \lambda) \;,
\end{align}
where $\lambda$ is discrete for bound states and continuous for scattering states, and
\begin{align}
E_{\mathbf{P},\lambda} = \frac{\mathbf{P}^2}{4m} + E_\lambda \;,
\end{align}
where $E_\lambda = E_\text{binding}$ for bound states and $E_\lambda = \mathbf{p}^2/m=mv^2/4$ for scattering states. $b_S$ is a destruction operator of the two-particle state in the Schrödinger picture. Inserting this into the correlator, we get in the energy basis the result:
\begin{align}
&G^{+-}_0(x,y;\mathbf{r},\mathbf{r}^\prime) = \\&\int\limits_{\mathbf{P},\mathbf{P}^\prime} e^{i \mathbf{P} \mathbf{x}} e^{-i \mathbf{P}^\prime \mathbf{y}}\sumint\limits_{\lambda, \lambda^\prime} \psi_\lambda(\mathbf{r}) \psi_{\lambda^\prime}^\star(\mathbf{r}^\prime) e^{-i E_{\mathbf{P},\lambda } (x_0 - t_0)} e^{i E_{\mathbf{P}^\prime,\lambda^\prime } (y_0 - t_0)} \text{Tr}[\rho(t_0) b^\dagger_S(\mathbf{P}^\prime, \lambda^\prime) b_S(\mathbf{P}, \lambda) ] \;. \nonumber
\end{align}
Now let us study the retarded correlator, which can also be written as:
\begin{align}
G^{R}_0(x,y;\mathbf{r}, \mathbf{r}^\prime)=\theta(x^0-y^0)[ G^{-+}_0(x,y;\mathbf{r}, \mathbf{r}^\prime) - G^{+-}_0(x,y;\mathbf{r}, \mathbf{r}^\prime) ] \;.   \label{eq:defretnew}
\end{align}
Using the canonical equal-time commutation relation,
\begin{align}
[b_S(\mathbf P, \lambda),b_S^\dagger(\mathbf P^\prime, \lambda^\prime)] = (2\pi)^3 \delta(\mathbf{P}-\mathbf{P}^\prime) \delta_{\lambda \lambda^\prime}, 
\end{align}
in \eqref{eq:defretnew}, one can directly see that $G^R$,
\begin{align}
G^{R}_0(x,y;\mathbf{r}, \mathbf{r}^\prime) = \theta(x^0-y^0) \int\limits_\mathbf{P} e^{-i \mathbf{P}(\mathbf{x}-\mathbf{y})} \sumint\limits_\lambda e^{-iE_{\mathbf{P},\lambda}(x^0-y^0)} \psi_\lambda(\mathbf{r}) \psi^\star_\lambda(\mathbf{r}^\prime) \label{eq:retfreeder} 
\end{align}
is time translational invariant (independent of $T$ and $t_0$). This implies that the advanced, and therefore also the spectral function are time translational invariant.

One can also see from \eqref{eq:retfreeder} that this solves the differential equations for the retarted correlator:
\begin{align}
(i \partial_{x^0} - h_x )G^{R}_0(x,y;\mathbf{r}, \mathbf{r}^\prime) &= i \delta^4(x-y) \delta(\mathbf{r}-\mathbf{r}^\prime) \\
(-i \partial_{y^0} - h_y^\prime )G^{R}_0(x,y;\mathbf{r}, \mathbf{r}^\prime) &= i \delta^4(x-y) \delta(\mathbf{r}-\mathbf{r}^\prime)
\end{align}
by using the completeness relation $\sumint_\lambda \psi_\lambda(\mathbf{r}) \psi^\star_\lambda(\mathbf{r}^\prime)=\delta(\mathbf{r}-\mathbf{r}^\prime)$. 

One can also start from the differential equations of $G^R_0$ to show that the solution is translational invariant in time. The retarded correlator satisfies the boundary condition:
\begin{align}
G^R_0(x,y;\mathbf{r},\mathbf{r}^\prime)=0 \text{ , for } x^0 > y^0. \label{eq:boundarycondret}
\end{align}
Translating the time coordinates,
\begin{align}
x^0 \rightarrow x^0 + \Delta \;, \quad y^0 \rightarrow y^0 + \Delta    
\end{align}
, and introducing 
\begin{align}
 G^R_{0,\Delta}(x,y;\mathbf{r}, \mathbf{r}^\prime)   \equiv G^R_0(x^0 + \Delta, y^0 + \Delta,\mathbf{x},\mathbf{y};\mathbf{r},\mathbf{r}^\prime) \;,
\end{align}
the differential equations become:
\begin{align}
(i \partial_{x^0} - h_x )G^{R}_{0,\Delta}(x,y;\mathbf{r}, \mathbf{r}^\prime) &= i \delta^4(x-y) \delta(\mathbf{r}-\mathbf{r}^\prime) \\
(-i \partial_{y^0} - h_y^\prime )G^{R}_{0,\Delta}(x,y;\mathbf{r}, \mathbf{r}^\prime) &= i \delta^4(x-y) \delta(\mathbf{r}-\mathbf{r}^\prime)
\end{align}
As the differential equation for $G^{R}_{0,\Delta}$ is the same as for $G^R_0$, and $G^{R}_{0,\Delta}$ satisfies the same boundary condition as $G^R_0$ given in \eqref{eq:boundarycondret}, it follows from uniqueness that they must be identical. Thus, $G^R_0$ is invariant under time translation. We will use this differential equation based method below, to show that full $G^R$ and $G^\rho$, for an equilibrium environment where $\Gamma=\Gamma(x-y)$, is invariant under time translation.

\subsection*{Initial density matrix assumptions}

Assuming a diagonal density matrix in momentum space,
\begin{align}
\text{Tr}[\rho(t_0) b^\dagger_S(\mathbf{P}^\prime, \lambda^\prime) b_S(\mathbf{P}, \lambda) ] = (2\pi)^3 \delta(\mathbf{P}-\mathbf{P}^\prime) \delta_{\lambda \lambda^\prime} f_\lambda(\mathbf{P}) \,,
\end{align}
the initial time $t_0$ dependence cancels and one arrives at:
\begin{align}
G_{0}^{+-}(t, \mathbf{R};\mathbf{r}, \mathbf{r}^\prime) = \int\limits_\mathbf{P} e^{-i \mathbf{P}(\mathbf{x}-\mathbf{y})} \sumint\limits_\lambda e^{-iE_{\mathbf{P},\lambda}(x^0-y^0)} \psi_\lambda(\mathbf{r}) \psi^\star_\lambda(\mathbf{r}^\prime) f_\lambda(\mathbf{P}) \;.
\end{align}
In this diagonal limit, one can recognize that the correlator only depends on the difference of time $t=x^0-y^0$ and difference of CM coordinates $\mathbf{R}=\mathbf{x}-\mathbf{y}$. This we used in Section~\ref{sec:integral}. Fourier transforming the difference variables one obtains:
\begin{align}
G_{0}^{+-}(\omega, \mathbf{K};\mathbf{r}, \mathbf{r}^\prime) = \sumint\limits_\lambda (2\pi) \delta(\omega - E_{\mathbf{K},\lambda}) \psi_\lambda(\mathbf{r}) \psi^\star_\lambda(\mathbf{r}^\prime) f_\lambda(\mathbf{K}) \;.
\end{align}
Assuming thermal equilibrium and a dilute system, 
\begin{align}
f_\lambda^\text{eq}(\mathbf{K})= e^{-\beta (2m+ E_{\mathbf{K},\lambda}) } \;,
\end{align}
 one can pull out the phase space distribution, leading to the same result when using the KMS condition of the $G_{\eta \xi,0}^{++--}$ correlator with zero chemical potential in the dilute approximation:
\begin{align}
G_{0}^{+-}(T,\omega, \mathbf{K};\mathbf{r}, \mathbf{r}^\prime)= e^{-\beta (2m +\omega)}\sumint\limits_\lambda (2\pi) \delta(\omega - E_{\mathbf{K},\lambda}) \psi_\lambda(\mathbf{r}) \psi^\star_\lambda(\mathbf{r}^\prime) \;,
\end{align}
where the temperature independent piece is the two-particle spectral function.

\subsection*{Translational invariance of $G^\rho$ and $G^R$}
We show that the spectral function in Section~\ref{sec:pnrderivation}, given by the solution of
\begin{align}
\left( i \partial_{x^0} - \frac{\mathbf{P}^2}{4m} + \frac{\Delta_\mathbf{r}}{m} - V(r)\right) G^\rho(x^0,y^0,\mathbf{P},\mathbf{r},\mathbf{r}^\prime) =-i \int_{y^0}^{x^0} \text{d}z^0 \delta(\mathbf{r}) \Gamma^\rho(x^0-z^0,\mathbf{P}) G^\rho(z^0,y^0,\mathbf{P};\mathbf{r},\mathbf{r}^\prime)\;, \nonumber 
\end{align}
is invariant under time translation. The spectral function obeys the boundary condition:
\begin{align}
G^\rho(x^0,x^0,\mathbf{P},\mathbf{r},\mathbf{r}^\prime) = \delta(\mathbf{r}-\mathbf{r}^\prime) \;.\label{eq:appgrhoboundary}
\end{align}
which follows from the canonical equal-time commutation relation of the two-body fields. We introduce a time translated spectral function as
\begin{align}
G^\rho_{\Delta}(x^0,y^0,\mathbf{P},\mathbf{r},\mathbf{r}^\prime) \equiv G^\rho(x^0+ \Delta ,y^0 + \Delta ,\mathbf{P},\mathbf{r},\mathbf{r}^\prime). 
\end{align}
Under time translation, the right hand side of the differential equation can be written as:
\begin{align}
&\int_{y^0+\Delta}^{x^0+\Delta} \text{d}z^0 \delta(\mathbf{r}) \Gamma^\rho(x^0+\Delta-z^0,\mathbf{P}) G^\rho(z^0,y^0+\Delta,\mathbf{P};\mathbf{r},\mathbf{r}^\prime) \\
&= \int_{y^0}^{x^0} \text{d}w^0 \delta(\mathbf{r}) \Gamma^\rho(x^0-w^0,\mathbf{P}) G^\rho(w^0+\Delta,y^0+\Delta,\mathbf{P};\mathbf{r},\mathbf{r}^\prime) \nonumber\\
&= \int_{y^0}^{x^0} \text{d}w^0 \delta(\mathbf{r}) \Gamma^\rho(x^0-w^0,\mathbf{P}) G^\rho_{\Delta}(w^0,y^0,\mathbf{P};\mathbf{r},\mathbf{r}^\prime) \nonumber \;,
\end{align}
where we changed the integration variable as $w^0=z^0-\Delta$. This shows that the differential equation for $G^{\rho}_{\Delta}$ is the same as for $G^{\rho}$. As $G^{\rho}_{\Delta}$ also satisfies the same boundary condition as $G^{\rho}$ given in \eqref{eq:appgrhoboundary}, it follows from uniqueness that they must be identical. Thus, $G^{\rho}$ is invariant under time translation for an equilibrium environment where $\Gamma=\Gamma(x-y)$. As the retarded correlator can be expressed through the spectral function via
\begin{align}
G^R(x^0,y^0,\mathbf{P},\mathbf{r},\mathbf{r}^\prime)=\theta(x^0-y^0)G^\rho(x^0,y^0,\mathbf{P},\mathbf{r},\mathbf{r}^\prime)\;,
\end{align}
it follows that the retarded correlator and hence also the advanced correlator are time translational invariant. This property was used in Section~\ref{sec:integral}.

\section{Technical details of the NREFT formalism}
\label{app:NREFT}
From our simplified NREFT action, which considers the particle number changing operator and a static potential only, we derive the equation for the four-point functions. We refer to it as a Bethe-Salpeter type of equation on the CTP contour. A formally exact Bethe-Salpeter equation can be derived directly from the 2PI effective action~\cite{vanHees:2001ik}, as 
\begin{equation}
\begin{split}
    G^{(4)} (x,y,z,w) =& \, G^{(2)} (x,z) G^{(2)} (y,w) \\
    &+ \int\limits_{\bar{x}^0, \bar{y}^0, \bar{z}^0, \bar{w}^0 \in \mathcal{C}} d^4 \bar{x} d^4 \bar{y} d^4 \bar{z} d^4 \bar{w} G^{(2)} (x,\bar{x}) G^{(2)} (y,\bar{y}) \Lambda (\bar{x}, \bar{y}, \bar{z}, \bar{w}) G^{(4)} (\bar{z}, \bar{w}, z, w).
\end{split}
\end{equation}
In this equation, $G^{(4)}$ is the full four-point function, $G^{(2)}$ is the full two-point function, and $\Lambda$ an interaction kernel. In principle, $\Lambda$ contains an infinite tower of diagrams to all loop orders, which we truncate at one-loop order. In this work, there are two relevant contributions entering $\Lambda$, the particle number changing term $\Gamma (x,y,z,w) = - \delta_\mathcal{C} (x,y) \delta_\mathcal{C} (z,w) \Gamma (x,z) $ and a long-range potential interaction term $D (x,y,z,w) = \delta_\mathcal{C} (x,z) \delta_\mathcal{C} (y,w) D (x,y)$. We then obtain for our correlator in Eq.~\eqref{eq:def4point}:
\begin{equation}
\begin{split}
    G_{\eta \xi} (x,y,z,w) =& \, G_\eta (x,z) G_\xi (y,w) + \int\limits_{\bar{x}^0, \bar{y}^0 \in \mathcal{C}}  d^4 \bar{x} d^4 \bar{y} G_\eta (x,\bar{x}) G_\xi (y,\bar{y}) D (\bar{x}, \bar{y}) G_{\eta \xi} (\bar{x}, \bar{y}, z, w) \\
    &- \int\limits_{\bar{x}^0, \bar{y}^0 \in \mathcal{C}}  d^4 \bar{x} d^4 \bar{y} G_\eta (x,\bar{x}) G_\xi (y,\bar{x}) \Gamma (\bar{x}, \bar{y}) G_{\eta \xi} (\bar{y}, \bar{y}, z, w).
    \label{eq:BS_full}
\end{split}
\end{equation}
Suppose the solution, $G_{\eta \xi, 0}$, to the equation with potential interaction only is known:
\begin{equation}
    G_{\eta \xi, 0} (x,y,z,w) = G_\eta (x,z) G_\xi (y,w) + \int\limits_{\bar{x}^0, \bar{y}^0 \in \mathcal{C}} d^4 \bar{x} d^4 \bar{y} G_\eta (x,\bar{x}) G_\xi (y,\bar{y}) D (\bar{x}, \bar{y}) G_{\eta \xi, 0} (\bar{x}, \bar{y}, z, w).
\end{equation}
In practice we assume that the potential $D$ is proportional to the time-contour delta function $\delta_\mathcal{C}(\bar{x}^0, \bar{y}^0)$, which defines what we mean by a static potential. This is also assumed in the pNREFT approach. It allows to reduce the problem to two-time correlators with four spatial arguments. Knowing the static potential solution, we can write for the full solution:
\begin{equation}
    G_{\eta \xi} (x,y,z,w) =G_{\eta \xi, 0} (x,y,z,w) - \int\limits_{\bar{x}^0, \bar{y}^0 \in \mathcal{C}}  d^4 \bar{x} d^4 \bar{y}  G_{\eta \xi, 0} (x,y,\bar{x},\bar{x}) \Gamma (\bar{x}, \bar{y}) G_{\eta \xi} (\bar{y}, \bar{y}, z, w).
    \label{eq:BS_Gamma}
\end{equation}
For our target component in the number density equation in \eqref{eq:number}, we need to evaluate the $++--$ component at equal spacetime points $x=y=z=w$. For a static potential, this quantity can be computed from the solution of $G_{\eta \xi}$ at  $t_1\equiv x^0=y^0$ and $t_2 \equiv z^0=w^0$. Thus we have to deal with correlators depending only on two time and four spatial coordinates. We introduce Wigner coordinates as:
\begin{align}
T=(t_1+t_2)/2\;, \quad t=t_1-t_2 \;, \quad \mathbf{X}_1 = (\mathbf{x} + \mathbf{y})/2 \;, \quad \mathbf{X}_2 = (\mathbf{z} + \mathbf{w})/2\;, \quad \mathbf{r} = \mathbf{x} - \mathbf{y} \;, \quad \mathbf{r}' = \mathbf{z} - \mathbf{w}.
\end{align}
We further define the average and relative center-of-mass coordinates
\begin{equation}
    \mathbf{X} = (\mathbf{X}_1 + \mathbf{X}_2)/2\;, \quad \mathbf{R} = \mathbf{X}_1 - \mathbf{X}_2.
\end{equation}
Throughout this work, we assume the system to be spatially homogeneous and isotropic, which implies in particular that the four-point function does not depend on $\mathbf{X}$ and we shall drop its dependence in the argument. The Wigner transformed four-point function is given by
\begin{align}
    G_{\eta \xi}^{\sigma_1 \sigma_2 \sigma_3 \sigma_4} (T, \omega, \mathbf{P} ; \mathbf{r}, \mathbf{r}') =& \int d t \, d^3 \mathbf{R} \, e^{i (\omega t - \mathbf{P} \cdot \mathbf{R})} \, G_{\eta \xi}^{\sigma_1 \sigma_2 \sigma_3 \sigma_4} \left(T, t, \mathbf{R}; \mathbf{r}, \mathbf{r}' \right), \nonumber \\
    G_{\eta \xi}^{\sigma_1 \sigma_2 \sigma_3 \sigma_4} \left(T, t, \mathbf{R}; \mathbf{r}, \mathbf{r}' \right) =& \int \frac{d \omega}{2 \pi} \frac{d^3 \mathbf{P}}{(2 \pi)^3} e^{- i (\omega t - \mathbf{P} \cdot \mathbf{R})} G_{\eta \xi}^{\sigma_1 \sigma_2 \sigma_3 \sigma_4} (T,\omega, \mathbf{P} ; \mathbf{r}, \mathbf{r}').
\end{align}
The Wigner transformed Bethe-Salpeter equation \eqref{eq:BS_Gamma} to leading order in the gradient expansion is given by
\begin{equation}
\begin{split}
    G_{\eta \xi}^{\sigma_1 \sigma_2 \sigma_3 \sigma_4} (T, P; \mathbf{r}, \mathbf{r}') =& \, G_{\eta \xi, 0}^{\sigma_1 \sigma_2 \sigma_3 \sigma_4} (T, P; \mathbf{r}, \mathbf{r}') \\
    &- \sum_{\sigma_5 \sigma_6} \sigma_5 \sigma_6 G_{\eta \xi, 0}^{\sigma_1 \sigma_2 \sigma_5 \sigma_5} (T, P; \mathbf{r}, 0) \Gamma^{\sigma_5 \sigma_6} (P) G_{\eta \xi}^{\sigma_6 \sigma_6 \sigma_3 \sigma_4} (T, P; 0, \mathbf{r}'),
\end{split}
\end{equation}
where $P = (P^0, \mathbf{P}) = (\omega, \mathbf{P})$. For $\sigma_1 = \sigma_2, \sigma_3 = \sigma_4$, and $\mathbf{r} = \mathbf{r}' = 0$, this simplifies to
\begin{equation}
\begin{split}
    G_{\eta \xi}^{\sigma_1 \sigma_1 \sigma_2 \sigma_2} (T, P; 0, 0) =& \, G_{\eta \xi, 0}^{\sigma_1 \sigma_1 \sigma_2 \sigma_2} (T, P; 0, 0) \\
    &- \sum_{\sigma_3 \sigma_4} \sigma_3 \sigma_4 G_{\eta \xi, 0}^{\sigma_1 \sigma_1 \sigma_3 \sigma_3} (T, P; 0, 0) \Gamma^{\sigma_3 \sigma_4} (P) G_{\eta \xi}^{\sigma_4 \sigma_4 \sigma_2 \sigma_2} (T, P; 0, 0).
\end{split}
    \label{eq:BS_static}
\end{equation}
This equation is the starting point for Section~\ref{sec:NRgradientall}.

\bibliographystyle{jhep}
\bibliography{main.bib}
\end{document}